\documentclass[conference]{IEEEtran}
\usepackage{algorithm}      % For captioned algos
\usepackage[noend]{algpseudocode}
\usepackage[inline]{enumitem}
\usepackage{multirow}
\usepackage{layouts}
\usepackage[table]{xcolor}
\usepackage{subcaption}
\usepackage{graphicx}
\usepackage{booktabs}
\usepackage{amsmath}
\usepackage{amsfonts}
\usepackage{amssymb}
\usepackage{tabularx}
\usepackage[T1]{fontenc}
\usepackage{url}
\usepackage[colorlinks,allcolors=blue]{hyperref}
\usepackage{fancyhdr}

\newcommand{\cmt}[1]{\ignorespaces}
\newcommand{\alt}[1]{\ignorespaces}

\newcommand{\refer}[1]{\ignorespaces}
\newcommand{\Context}[1]{\ignorespaces}
\newcommand{\Gap}[1]{\ignorespaces}
\newcommand{\Innovation}[1]{\ignorespaces}
\newcommand{\cbd}[1]{\ignorespaces} % {\textcolor{yellow}{#1}}  % cbd = can be deleted

\newcommand{\system}{FemmIR} %{WeSProS} %{FemmIR} 
\newcommand{\setkp}{Q_H}
\newcommand{\kp}{q_H}
\newcommand{\datatext}{InciText} %{\system-text}

\newcommand{\mmirdataexp}{MuQNOL}%{(\datatext~+ MARS)~}
\newcommand{\simhart}{SIM}
\newcommand{\textattrname}{\textsc}
\newcommand{\textstring}{\texttt}
\newcommand{\textattrvalue}{\texttt}
\algnewcommand\repropvalues[1]{\textsc{RegexPropVal(#1)}}
\newcommand{\posalgo}{\textsc{POSID}} 

\newcommand{\harg}{\textsc{HARG}} 
\newcommand{\har}{\textsc{HAR}} 

\newcommand{\querygmath}{g^q}
\newcommand{\comparisongmath}{g^c}

\newcommand{\QquerydatasamplefromAnyModality}{$\mathbf{d}^{m}_{q}$}%{$Q_{\mathbf{d}^{m \in \mathcal{M}}}$}
\newcommand{\querydatasamplefromModalitym}{$\mathbf{d}^{m}_{q}$}
\newcommand{\dataInResultSet}{$\mathbf{d}^{x_c}_{c}$}
\newcommand{\bigOP}{\mathcal{O}} % bigObjectProp
\newcommand{\smallOP}{\mathbf{o}} % smallObjectProp
\newcommand\mycommfont[1]{\scriptsize\textcolor{gray}{#1}} % \footnotesize\ttfamily
\algnewcommand\rcost{\textsc{rcost($\smallOP_p$)}}
\algnewcommand\icost{\textsc{icost($\smallOP_p$)}}
\algnewcommand\listPropED[1]{\textsc{\propled}(#1)}%{Prop-List-Edit-Distance}
\algnewcommand\listPropComp[1]{\textsc{\hashcmp}(#1)}%{Hashmap-Compare}
\algnewcommand\length{\textsc{LENGTH}}
\algnewcommand\type{\textsc{TYPE}}
\algnewcommand\simgnn{\textsc{SIMGNN}}
\newcommand{\ced}{CED}
\algnewcommand\ced{\textsc{\cedoutsidealgo}}
\algnewcommand\CostMat{\textsc{C}}
\algnewcommand\eplvcost{\textsc{$\CostMat_{i, j}$}}
\algnewcommand\ordcmp{\textsc{oComp}} 
\algnewcommand\munkersType{\textsc{mType}}
\algnewcommand\zpinq{\textsc{$z_p(u_i)$}} % {z^q_p}
\algnewcommand\zpinc{\textsc{$z_p(v_j)$}}
\algnewcommand\algorithmicforeach{\textbf{foreach}}
\algdef{S}[FOR]{ForEach}[1]{\algorithmicforeach\ #1\ \algorithmicdo}

\algrenewcommand\algorithmicrequire{\textbf{Input:}}
\algrenewcommand\algorithmicensure{\textbf{Output:}}

\algnewcommand\mynewlinecomment[1]{
\newline \null\hfill\algorithmiccomment{{\mycommfont{#1}}}
}
\algnewcommand\mysamelinecomment[1]{
\algorithmiccomment{{\mycommfont{#1}}}
}

\newtheorem{example}{Example}%[section]
\newtheorem{problem}{Problem}[section]
\newtheorem{definition}{Definition}[section]

\begin{document}
% \title{Bare Demo of IEEEtran.cls\\ for IEEE Conferences}

% \title{Feature Centric Multi-modal Information Retrieval in Open World Environment}
\title{Multimodal Information Retrieval for Open World with Edit Distance Weak Supervision}  %(\system)}
% \title{Multi-modal Information Retrieval with Weak Supervision from Object-Property Similarity}
% \title{Multi-modal Information Retrieval with Edit Distance based on Object-Property Similarities (\system)}


\author{\IEEEauthorblockN{KMA Solaiman}
\IEEEauthorblockA{Department of Computer Science\\
Purdue University\\
West Lafayette, IN 47906, USA\\
Email: ksolaima@purdue.edu}
\and
\IEEEauthorblockN{Bharat Bhargava}
\IEEEauthorblockA{Department of Computer Science\\
Purdue University\\
West Lafayette, IN 47906, USA\\
Email: bbshail@purdue.edu}
% \and
% \IEEEauthorblockN{James Kirk\\ and Montgomery Scott}
% \IEEEauthorblockA{Starfleet Academy\\
% San Francisco, California 96678--2391\\
% Telephone: (800) 555--1212\\
}


%Telephone: (800) 555--1212, Fax: (888) 555--1212}


\maketitle

% footnote
% First-page footnote after maketitle
\begingroup\renewcommand\thefootnote{\textit{}}\footnotetext{
This paper was originally submitted to ICDE 2024 (Nov 2023). An earlier version of this paper appeared on TechRxiv: DOI: 10.36227/techrxiv.21990284.v1, uploaded on February 05, 2023.
}\endgroup

% \thispagestyle{plain} % TODO: omit
% \pagestyle{plain}     % TODO: omit

\begin{abstract}
% problem of before
Existing multi-media retrieval models either rely on creating a common subspace with modality-specific representation models or require schema mapping among modalities to measure similarities among multi-media data. 
% The heterogeneity gap between explicitly mentioned properties in the information needs and low-level representation features used in the retrieval models makes them unusable in certain systems. 
% what is our goal
Our goal is to avoid the annotation overhead incurred from considering retrieval as a supervised classification task and re-use the pre-trained encoders in large language models and vision tasks. 
% Our intuition is we can introduce weak supervision for relevance from the similarities among the object properties described in a sample.
% what do we propose
We propose ``\system'', a framework to retrieve multimodal results relevant to information needs expressed with multimodal queries by example without any similarity label.
% Why is it necessary
Such identification is necessary for real-world applications where data annotations are scarce and satisfactory performance is required without fine-tuning with a common framework across applications. We curate a new dataset
called \mmirdataexp\ for benchmarking progress on this task.
% What is the technique
Our technique is based on \textit{weak supervision} introduced through \textit{edit distance} between samples: graph edit distance can be modified to consider the cost of replacing a data sample in terms of its properties, and relevance can be measured through the implicit signal from the amount of edit cost among the objects.
% How is it different
Unlike metric learning or encoding networks, \system\ re-uses the high-level properties and maintains the property-value and relationship constraints with a multi-level interaction score between data samples and the query example provided by the user.
We also proposed a novel attribute recognition model from unstructured text  ``HART'' that can identify attributes without finetuning or large language models.
% how do we evaluate it
We empirically evaluate \system\ and HART on a missing person use-case with \mmirdataexp.
% how does it perform
HART successfully identifies human attributes from large unstructured text without additional training, while
\system\ performs comparably to similar retrieval systems in delivering on-demand retrieval results with exact and approximate similarities while using the existing property identifiers in the system. 
\end{abstract}


% \IfCLASSOPTIONpeerreview
% \fi
\IEEEpeerreviewmaketitle

\section{Introduction}
% This demo file is intended to serve as a ``starter file''
% for IEEE conference papers produced under \LaTeX\ using
% IEEEtran.cls version 1.8b and later.
% % You must have at least 2 lines in the paragraph with the drop letter
% % (should never be an issue)
% I wish you the best of success.


With the influx of multimedia data sources, 
% exploratory data analysis has become increasingly challenging in modern applications. 
comparing data from different modalities to grasp a more informed decision for any phenomenon has become increasingly difficult.
Humans analyze information across modalities using many indirect cues and common hints. But when it transfers to hours of videos or thousands of documents, it is imperative to have a recommender system that filters out the most important information according to the user's preference and recommends with an unseen item that this user is likely to interact in the future. Existing sequential recommenders propose contrastive learning \cite{luo2023end} or attention networks to do representation alignments between multiple modalities based on similarity labels across data from different modalities. Without the similarity labels, it is impossible to adapt these retrieval models by fine-tuning them to certain applications, which is the most common scenario for most real-world use cases. 
% With the ever-growing size of the multi-media data, multi-modal data analysis becomes difficult for any real-world application-specific information needs, 

% Challenges for real-world environment:\\
% 1. No Similarity/ Relevance label\\
% 2. Large amount of data (show ingestion from SKOD)\\
% 3. External Features extractors are available for different modalities, but they have mismatched features/ properties

\begin{example}[Application-specific Information Need] %[Approximate Similarity]
% [Google result - wikipedia data, similarity in between document and a still pic, or the knowledge graph in google result (not multi-level similarity)]
\label{example:obj-prop-focus-info-need} %{example:approx-sim}
\textit{
% A law enforcement agency 
An organization wants to build an automated system to find video frames containing persons of interest from many hours of video feeds, connect them to previous occurrences from incident reports, and find patterns among these occurrences.  
% text queries were considered to be text modalities. 
Alex is asked to develop a machine learning (ML) pipeline over these datasets to predict the videos where the person mentioned in the text would be found, and, subsequently, the authority would look for them in those videos. Alex decides to use an off-the-shelf retrieval algorithm that is trained over video and text. However, the performance was not satisfactory, as: (1) Alex could not modify the model to focus on specific properties that are most common for a missing person as he did not know which video frames contained the person mentioned in the reports, and (2) he could not perform transfer learning as the annotated data is very difficult to achieve in this case, where one positive case occurs in 8-10 hours of video.
Now he wonders: (1) how can he re-train the retrieval model without any training data to focus similarity on the desired properties? (2) if he runs a property identifier in each data modality and performs only explicit matching would that achieve the desired performance? (3) how can he map similar properties that are described differently from each modality?
% not just gender and race match - cant have zero and one, there can be partial matching.
% Vitivr matches low level features.
}
\end{example}

% \begin{figure*}[h]
%   \centering
%   \begin{subfigure}{0.2\textwidth}
%     \centering
%          \includegraphics[
%          trim={0 5cm 0 4.1cm},clip, width=1.2\textwidth, height=1.1\linewidth]{images/mono_county.png}
%         \caption{Flyers}
%         \label{fig:upload-flyers}
%      \end{subfigure}%
%      \begin{subfigure}{0.2\textwidth}
%          \centering
%           \includegraphics[ width=0.6\textwidth, height=1.1\linewidth]{images/0100C1T0006F112.jpg}
%           \caption{Screenshots}
%           \label{fig:upload-screenshot}
%      \end{subfigure}
%      \caption{Upload Example of Data Objects}
%     \label{fig:upload-example}
%     \begin{subfigure}{0.2\textwidth}
%          \centering
%           \includegraphics[ width=0.6\textwidth, height=1.1\linewidth]{images/0100C1T0006F112.jpg}
%           \caption{Screenshots}
%           \label{fig:upload-screenshot}
%      \end{subfigure}
%      \caption{Upload Example of Data Objects}
%     \label{fig:upload-example}
% \end{figure*}

To address these issues, we propose a \textbf{Fe}ature-centric \textbf{M}ulti\textbf{m}odal \textbf{I}nformation \textbf{R}etrieval model for open-world environment, \textbf{FemmIR}. Our framework designs multiple plug-and-play components with effective representation alignment and matching objectives to enable ranked information retrieval across application domains and modalities. Specifically, we leverage pre-trained text and vision property identifiers as feature extractors.
% to extract the modality-specific feature embeddings
The modality-specific high-level features are fused into multimodal item representation via a graph representation approach and an attention network, which is subsequently processed by a graph similarity approximation model to capture the implicit similarities. Since we assume no similarity label is available across multiple modalities, we needed weak supervision from other sources \cite{li2020weakly, mithun2019weakly, alaudah2019structureweakly, solaiman2022weakly}. We hypothesize that capturing how much change is needed to convert a data sample to another can provide us with a source of weak supervision. Considering a data sample as a collection of objects with certain properties along with the relationships among them, we modeled a novel distance metric based on graph edit distance (GED).
% Since multi-media data is usually large in number, it is expensive to annotate the relevance for each different system. 
% Prior works in retrieval system \cite{li2020weakly, mithun2019weakly, alaudah2019structureweakly, solaiman2022weakly} use inexact weak supervision to tackle this issue of annotation expense.
% To achieve weak supervision, we propose a novel distance metric. 

% releated work: with the advent of attention network and large language models people have learned to align multmodal data very well. But all the existing model require relevance label and has been benchmarked on well-known and clean multimedia dataset - XMedia, Wiki, NUS. We are working on real world data where there is no relevance label.

% EARS does not use relevance label but is slow compared to embedding models which can infer from pre-trained model.s

% MuQNOL(\textbf{Mu}ltimodal \textbf{Q}ueries with \textbf{NO} similarity \textbf{L}abel)

We introduce a new benchmark and dataset
called \mmirdataexp\ (\textbf{Mu}ltimodal \textbf{Q}ueries with \textbf{NO} similarity \textbf{L}abel) to train and evaluate models to retrieve the relevant data from a multimodal corpus given multimodal (vision + language) queries without any similarity labels. To create this dataset, we start with the MARS \cite{zheng2016mars} dataset as a source – 
% MARS contains pedestrian images annotated with pedestrian attributes.
MARS is a large-scale dataset of pedestrian image sequences with 16 annotated attributes. 
% MARS consists of 20,478 tracklets from 1,261 people captured by six cameras.
We combined it with an unstructured text dataset, \datatext, consisting of incident reports, press releases, synthetic reports, and officer narratives from old police cases. We also annotated \datatext\ \cite{solaiman2023feature} for three attributes with a wide range of possible values. We proposed a novel \textsc{\textbf{H}}uman \textsc{\textbf{A}}ttribute \textsc{\textbf{R}}ecognition model from unstructured \textsc{\textbf{T}ext}, HART to identify these attributes from \datatext.
% It contains 13 press releases, 40 officer narratives, and five incident reports.
% How?
From the 16 attributes in MARS, we select common attributes in MARS and \datatext\ for \mmirdataexp\ where the retrieved answer includes both an image, video, and text.

Unlike existing multimodal retrieval models, \system\ does not require a large amount of training data, and data representations can be aligned through the weak supervision. Among existing works, correlation learning methods \cite{rasiwasia2010CCA, rupnik2010multiMCCA, zhang2017generalized, wang2015deep, kan2016multi, peng2017ccl, peng2016cmdn} linearly or non-linearly projects low-level features from representation models to a common subspace. Metric learning methods \cite{faghri2017vse++, xu2019deep, wei2020universal} learn a distance function over data objects based on a loss function to map them into the common subspace. \system\ closely relates to metric learning methods. Contrary to them, we do not directly correlate class labels or weak labels to the loss function. The proposed edit distance between property graphs implicitly captures the signal for relevance.
% \cite{hu2019deep, Wang2021DRSLDR} builds modality-independent learning models.  
%
% Late fusion and entity matching is related to our problem. \cite{dong2014dataprism/10.14778/2732951.2732962, wang2015dataxrayDataprism/10.1145/2723372.2750549}. DICE \cite{DICE-elkindi/10.14778/3476311.3476353} finds data sources, 3 good things - soft sim score, using existing prop, no need to map if similarity known for training data.
%
In contrast to common representation learning models, \textit{data discovery} models based on relational queries allow more flexibility to consider explicit information needs from users, and use high-level properties in the system. EARS \cite{solaiman2022computer} is one such content-based data discovery system that, similar to our approach, takes user examples as queries and delivers relevant multi-media results. However, the prime aspect of EARS is it assumes a schema mapping among all modalities, the number of JOIN queries increases as a product of the number of properties-of-interest and modalities, and to introduce new modalities the common schema needs to be updated.
In contrast, \system\ offers a general solution to include retrieval from novel modalities for a diverse set of systems and does not need an explicit design for each new modality. 

Our contribution and findings are listed below.
\begin{itemize}
    \item We introduce a new dataset \mmirdataexp\ to facilitate research on multimodal information retrieval for real-world use cases without similarity labels.
    \item We propose an end-to-end retrieval model, \system, that delivers ranked results from multimodal data relevant to a given multimodal query using weak supervision from a novel distance metric \ced. %\system\ does not assume
    \system\ does not need any similarity label and can be pre-trained on any application domain for faster inference time.
    \item We benchmarked another end-to-end model, EARS on \mmirdataexp\ dataset. EARS is an exact inference model for multimodal data retrieval. We observed that \system\ is a capable multimodal retriever that surpasses existing multimodal knowledge retrieval methods without fine-tuning.
    \item 
    % We proposed a novel object attribute detection model, HART. 
    We proposed HART, to retrieve attribute values from the unstructured text as part of the basic property identifiers for the \system\ framework. 
\end{itemize}

\section{Preliminaries \& Problem Definitions}
In this section, we first provide formal definitions relevant to our proposed methods. We then proceed to formulate the problem of multimodal information retrieval for unconstrained data for property-specific information needs, and the problem of object-property identification from text.

% https://ieeexplore.ieee.org/stamp/stamp.jsp?tp=&arnumber=6313232
\begin{definition}[Attributed relational graph]
\label{defn:arg}
An attributed relational graph (ARG) is a graph whose nodes and edges have assigned attributes (single value or vectors of values). 
% For the sake of simplicity, we denote the node and edge attributes by labels. %, as labels are specific types of attributes. 
Although we focus our methodology only on directed and labeled graphs, it is designed to handle any form of graphs. 
An ARG is defined as:
$g = (N, E, l)$
where
\begin{enumerate}
    \item $N$ is the finite set of nodes, 
    \item $E \subseteq N \times N$ is the set of edges,
    \item  $l : N(g) \cup E(g) \rightarrow \Sigma$ is a labeling function that assigns labels to each vertex and edge from $\Sigma$. 
    % Specifically, $l(u)$ and $l(u, u')$ are the label of node $u$ and the label of edge $(u, u')$, respectively,
    \item $\Sigma$ is a % finite or infinite 
    set of unconstrained labels. $A \in \Sigma$ represents labels enumerating the node-type.
\end{enumerate}

\end{definition}
\cbd{
\begin{definition} [Graph Edit Distance]
\label{defn:ged}
Formally, the edit distance between $g_1$ and $g_2$, denoted by
GED ($g_1, g_2$), is the number of edit operations in the optimal alignments that transform $g_1$ into $g_2$, where an edit operation on a graph $g$ is an insertion or deletion of a node/edge or relabelling of a node/edge. Intuitively, if two graphs are identical (isomorphic), their GED is 0. 
\end{definition}
\begin{definition}[Parts-of-Speech]
\label{defn:pos}
Parts-of-Speech (POS) is a category to which a word or phrase is assigned in accordance with its syntactic functions. 
A syntactic function is the grammatical relationship \footnote{https://glossary.sil.org/term/grammatical-relation} (i.e., subject, object) of one constituent to another within a syntactic construction.
\end{definition}
}
\begin{definition}[Wu-Palmer Distance]%[Wordnet Synsets]
\label{defn:wordnets}
Wordnet \cite{wordnet98} is a lexical knowledge base where words are organized in a hypernym tree based on their origin. 
% Words are grouped into Synsets based on their synonyms.
Wu-Palmer distance calculates the similarity between word meanings based on the similarity between the word senses and the location of the synsets relative to each other in the hypernym tree. 
Given the synsets of two strings $s_{t_1}$ and $s_{t_2}$, and the LCS (Least Common Subsumer) between them, the Wu-Palmer distance is:
\begin{equation}
    wpdist(s_{t_1}, s_{t_2}) = 2 * \frac{depth(lcs(s_{t_1}, s_{t_2}))}{depth(s_{t_1}) + depth(s_{t_2})}
\end{equation}
\end{definition}
\begin{definition}[Natural Language Inference]
\label{defn:nli}
Given a hypothesis $h$ and a premise $p$, 
natural language inference (NLI) is the task of determining the probability $Pr$ of the hypothesis being true (entailment $E$), false (contradiction $C$), or undetermined (neutral $N$). NLI determines the best label $l$: 
\begin{equation} \arg\max_{l \in \{E, C, N\}} Pr(l ~|~ h, p) \end{equation}
% the task of determining whether a "hypothesis" is true (entailment), false (contradiction), or undetermined (neutral) given a "premise".
\end{definition}

%%%%%%%%%%%%%%%%%%%%%%%%%%%%%%%%%%%%%%%%%%%%%%%%%%%%%%%%%%%%%%%%%%%%%%%%%%%%%%%%%%%%%%%%%%%%%%%%%%%%%%%%%%%%%%%
\subsection{Problem Definitions}
%%%%%%% https://epubs.siam.org/doi/epdf/10.1137/20M1386062 -- defines dynamic graph 3.3.1
%%%%%  https://ieeexplore.ieee.org/stamp/stamp.jsp?tp=&arnumber=8294302 -- defines heteregenerous grpah 3.1.2 
%
% vitrivr supports several query modes, including Query-by-Sketch
% (QbS), Query-by-Example (QbE), relevance feedback, and textual
% queries. These query modes can be combined within a single query
% and can span several media modalities.
%
% {\Huge \textbf{$d^{modality}_{j-th}$}}
% d/D - i, j, n, q, x, x_q, 
% o - l, k, p, z, z_r
% modality - m
%
Assuming a collection of data from $\mathcal{M} \in \mathbb{Z}^{+}$ modalities, %$\mathcal{D}$, 
we denote the set containing $n_i \in \mathbb{Z}^{+}$ samples from the $i$-th modality as 
$
\mathcal{D}_i = \{ \mathbf{d}^i_1, \mathbf{d}^i_2, \ldots, \mathbf{d}^i_{n_i} \},
$
where $j$-th sample of the $i$-th modality is $\mathbf{d}^i_j$. %, and 
% Each data sample contains a collection of \textit{object-properties}. For example, a document has a \textit{topic, metadata}, and \textit{entities} with their \textit{relationships}, along with any \textit{event} it describes.
% (Table \ref{table:object-properties}). 
%
% Let $\bigOP = 

Any data sample $\mathbf{d}^i_j$ is described with a subset of object properties, $\bigOP = \{\smallOP_1, \smallOP_2, \ldots, \smallOP_{l}\}$ 
% be the set of all such \textit{object-properties}, 
where $z_r$ is the set of \cmt{categorical} values of $\smallOP_r$. 
% $\bigOP_E \subseteq \bigOP$ denotes the set of object-properties describing an entity $E$.
% $z_r = \{\phi\}$ indicates that $\smallOP_r$ is not present in \alt{the data sample} $\mathbf{d}^i_j$.
%
Property identifiers implement a relation, \textit{PROP} ($\mathbf{d}^i_j$) $\subset \bigOP$
that maps a data-sample to a set of object-properties.
% \textit{PROP}: $\mathcal{D} \rightarrow \bigOP$. 
%
A query is issued against a corpus with $\mathcal{M}$-modalities, $\mathcal{D} = \{ \mathcal{D}_1, \mathcal{D}_2, \ldots, \mathcal{D_M} \}$.\\

\begin{problem} [Multimodal Information Retrieval]
\label{problem:mmir}
Given a query in modality $m$, \QquerydatasamplefromAnyModality, 
% = $\{\smallOP_1 = z_1, \smallOP_2 = z_2, \ldots, \smallOP_{p} = z_p\}$, 
the task is to retrieve a ranked list, 
$R = (\mathbf{d}^{x_1}_{1}, \mathbf{d}^{x_2}_{2}, \dots \mathbf{d}^{x_t}_{t}) $ 
of $t \in \mathbb{N}_0$ data-samples from all available modalities in the system satisfying
\textit{PROP} (\querydatasamplefromModalitym).
% where \dataInResultSet~ is $c$-th data $\in R$ from modality $x_c \in_R \mathcal{M}$.
% is a data query that is expressed in one of the two ways: 
\QquerydatasamplefromAnyModality~ can be expressed in two different ways: 
\begin{enumerate}[label=(\textbf{\arabic*})]
    % [Query-by-Properties] 
    \item (Query-by-Properties) \querydatasamplefromModalitym~ with $p$ object-properties $\{\smallOP_1 = z_1, \smallOP_2 = z_2, \ldots, \smallOP_{p} = z_p\}$,
    or
    \item (Query-by-Example) An example data-sample, \querydatasamplefromModalitym~ with the \textit{PROP} relation. % (\querydatasamplefromModalitym). 
    % = \QquerydatasamplefromAnyModality.
\end{enumerate}
\end{problem}
%
%
% ALTERNATIVE: 
Relevance is scored \alt{calculated} based on the degree of common object-properties between a data-object 
\dataInResultSet~ in the ranked list, and the query data \querydatasamplefromModalitym,
\textit{PROP} (\querydatasamplefromModalitym) 
$\cap$
\textit{PROP} (\dataInResultSet). A similarity score is used to define the degree of relevance, \[0 \leq \text{SIM (\dataInResultSet, ~\querydatasamplefromModalitym)} \leq 1. \]
Similarity score of 0 indicates non-relevance, whereas a score of 1 indicates complete relevance and a proper subset,\\
\textit{PROP} (\querydatasamplefromModalitym) 
$\subset$
\textit{PROP} (\dataInResultSet).
%
%

% Assumptions in the problem setting - 
% Our problem setting assumes that the user has knowledge about $o_p$ and their corresponding $z_p$ for Query-by-Properties. This assumption is realistic in real-world scenarios and has been considered in 
% % based on
% multimodal data query literature where properties are used to express the information need \cite{solaiman2022computer, Vitrivr/10.1145/3323873.3326921, QbE/Event/10.1145/3397271.3401283}.

% EXTRA
% Certain applications allow access to multiple examples for Query-by-Example. \system\ can be extended to this setting as well by first using \textsc{dataprism} \cite{Dataprism-annafariha/10.1145/3514221.3517864} to retrieve user information needs.

% begin of hart
\subsubsection{Property Identification from Unstructured Text} 
As a relevant sub-task, we explore the problem of identifying properties describing humans from unstructured text. As discussed in SurvQ \cite{HILDA2020}, a finite number of visible and approximate properties such as, \textattrname{gender, race, build, height, clothes}, etc. are used in describing a person-of-interest to search for them. We denote these properties for person identification as $\bigOP_H$.
\begin{example}
\label{example:person-profiling}
%For example, 
The sentence ``\textstring{a ${}^\dag$person with white ethnicity and \textbf{medium} build was seen in Vernon St., \textit{wearing} \textbf{white jeans} and \textbf{blue shirt}}'' describes properties of a person: 
\begin{enumerate}
    % \item \textattrname{gender} = \textattrvalue{male},
    % \item \textattrname{race} = \textattrvalue{white},
    \item \textattrname{build} = \textattrvalue{medium},
    \item \textattrname{${}^*$clothes} = \{\textattrvalue{jeans, shirt}\},
    \item \textattrname{upper-wear-color}= \{\textattrvalue{white}\}, 
    \item \textattrname{bottom-wear-color} = \{\textattrvalue{blue}\}, and 
    \item \textattrname{relation} = \{\textattrvalue{wearing, ${}^\dag$Person, ${}^*$Clothes}\}.
\end{enumerate}
\end{example}
% TODO: double-check if there is any assumption for this. Examples were mentioned/ DONE

\begin{problem}[Human Attribute Recognition from Text]
\label{problem:hart}
% Don't wanna use D to denote text document, as D is for data-sample
% vuila gesi
Given a large text $T$ with $T_s$ sentences, each with $|w|$ tokens, the problem of human attribute recognition from $T$ is to
\begin{enumerate}
    \item identify the 
    % text snippet $T_s \subset T$ 
    set of sentences~ $C_s \subset T_s$ 
    that describes properties of a person,
    \item expose the set of object-properties \alt{names} ~$\bigOP_H$ %$o_p \subset \mathcal{O}_H$ 
    from $C_s$ and 
    \item \cmt{Upon identifying $o_p$,} extract the set of values $z_p$ of the identified properties $\smallOP_p$. 
    % can combine 2 and 3 if needed
\end{enumerate}
\end{problem}
%
%
%
% the assumptions
Our problem setting assumes that the set of key-phrases ($\setkp$) often used in 
% person profiling 
sentences describing properties of a person % keeps continuation from Problem 2.2 
are either known (provided by domain experts), or a small amount of annotated documents are provided to identify $\setkp$~ manually. In Example \ref{example:person-profiling}, $\setkp$~ = \{\textstring{wearing}\}. % , shirt
% TODO: add citations
The first assumption is derived from the literature on pedestrian attribute recognition from visual and textual modalities, and the second assumption is computationally inexpensive.
% holds as small amount of curated % non-annotated 
% data is always available for a problem setting. 
Note that, ($\setkp ~\cap~ \bigOP_H) \ne \{\phi\}$.
% Candidate sentences are sentences in the text that mentions phrases similar to the key-phrases within an empirical threshold value.
\begin{definition} [Candidate Sentences]
Given a collection of sentences $T_s$, key-phrase for describing an object in text $\kp ~\subset~ \setkp$, and an empirical threshold $\theta_H$, Candidate sentence is
\begin{equation}
\label{eqn:cand-sent-def}
    C_s = \{s: s \in T_s, \kp \in \setkp ~|~ \simhart(\kp, s) > \theta_H \}
\end{equation}
\end{definition}
%%%%%%%%%%%%%%%%%%%%%%%%%%%%%%%%%%%%%%%%%%%%%%%%%%%%%%%%%%%%%%%%%%%%%%%%%
%

%%%%%%%%%%%%%%%%%%%%%%%%%%%%%%%%%%%%%%%%%%%%%%%%%%%%%%%%%%%%%%%%%%%%%%%%%
\section{MULTIMODAL INFORMATION RETRIVAL}
\label{sec:mmir}

% We will now describe the multimodal similarity matching method to find the relevant data to user provided information need (mentioned with an example, or with object properties). 
Our proposed similarity matching algorithm considers the data samples from the data repository or the data streams, \dataInResultSet~ and user-provided example, \querydatasamplefromModalitym~ as input and outputs the similarity score between them: $SIM$ (\dataInResultSet, \querydatasamplefromModalitym). The corresponding object properties are assumed to be extracted by the system-specific property identifiers in the pre-processing stage. We propose a novel distance metric to rank the data samples based on the number of edits needed to convert the properties of one sample to another instead of manually annotating or aggregating the number of matched object properties. To this end, we first process the extracted properties from the input data samples with a graph encoding mechanism which converts the properties into a hierarchical attributed relational graph (\harg) and generates a graph representation for each sample. Then \system\ adopted the Munkers' algorithm \cite{riesen2007bipartite} to calculate the proposed edit distance between the data samples and use it as a similarity label. Finally, we used an edit distance approximation algorithm with Neural Tensor Network (NTN) to learn a function to map the graph embedding of the {\harg s} to a similarity score between the data samples. During inference, the model just takes the extracted properties from the data samples and outputs the similarity score by using the mapping function. We start with a use case to demonstrate the information retrieval task and our observations that led to the proposed system.
% \system\ works and then proceed to describe the algorithm.

%
% % \input{sections/graph.tex}
%
% % \input{sections/weak-label-generation.tex}
%
% % \input{sections/feemir-algo.tex}
%
% \subsection{Data Ingestion with Graphs}
\subsection{Graph Representation for Data Sample}
\label{subsec:data-ingestion-w-graphs}
Consider the task of finding the location of a person from a large amount of video data using text queries or reports.
% (Example \ref{example:obj-prop-focus-info-need}). 
The system finds the video feeds that have persons similar to the report description 
% (using multi-modal similarity matching) 
by focusing on the object properties of the person in the video and text. 
The goal is to identify the similarity score between video feeds, text queries, and incident reports which can then be used to deliver a ranked list of relevant data samples to the user. 
\begin{figure*}[htb]
    \centering
    \includegraphics[width=\textwidth]{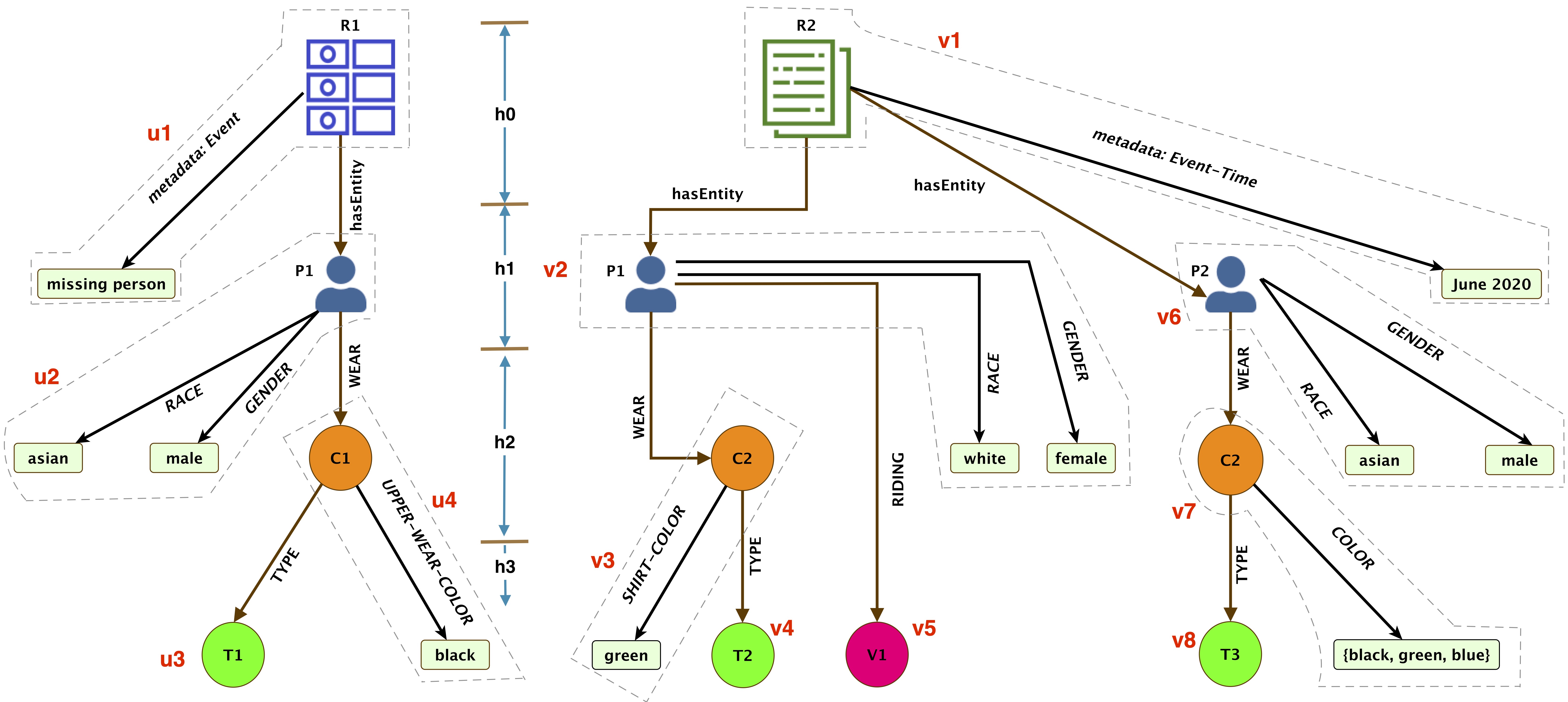}
    \caption{
    HARG and Weak Label Generation; Left sided graph refers to $\querygmath$, and the right sided graph refers to $\comparisongmath$. 
    Node-type labels are as follows. 
    V: EPL Vertex,
    R: Root,
    P: Person,
    C: Clothes,
    T: Type,
    M: Motor-Vehicles. 
    Squared nodes correspond to the non-empty leaf nodes.
    }
    \label{fig:qg-and-cg}
\end{figure*}
%
%
% \textcolor{pink}{
% In Example~\ref{example:obj-prop-focus-info-need}, user specified the object-properties to identify a person such as, \textattrname{gender, race, cloth descriptions, place, time-frame}, etc.
% For the retrieval matching, we only focused on those object-properties. Figure~\ref{fig:qg-and-cg} shows the graphs that we built from these properties. One of them refers to the query example user provides, another is for the data sample from the data repository, or the data stream.
% }
%
% \paragraph{Observations} 
We make the following observations for property-specific multimodal queries:
\begin{enumerate}[label=\textbf{O\thesection.\arabic*}, topsep=0pt, leftmargin=*, series=obsmmir]%, leftmargin=2 pt, [series=theoremconditions]
\item The number of object properties between two data samples is finite, and the values of the properties are mostly categorical values. 
\item A data sample can describe a large number of objects and object properties, but for system-specific similarity comparison, a user is only interested in a finite number of properties.
\item Data samples are special objects with different properties such as metadata, topics, and events. 
% \textit{Entities} are specific types of objects described in a data sample with their own properties.
\item \label{itm:mmir-relationship} \textit{Relationships between objects} are specific types of object properties that belong to all participating objects. The set of values corresponding to the objects would be complementary to each other.
Value for relation-name can be different for the same relationship through different data samples. For example, different text would describe the same action in different forms: \textit{wearing, wear, has}. 
\item \label{itm:mmir-single-multi-value-zp} Some properties in $z_p$ have single and fixed value-set i.e., \textattrname{gender, race, height}, while other properties have multiple values in their value-set i.e., \textattrname{clothes}.
\item \label{itm:mmir-different-valued-zp}
% The multi-value 
Some object properties such as, \textattrname{clothes-color} have different values for different data samples. For example, in Figure~\ref{fig:qg-and-cg}, \textattrvalue{UPPER-WEAR-COLOR,  SHIRT-COLOR, COLOR} all refer to the color of clothes. 
\end{enumerate}
 
Our intuition here is that entities, relationships, and object properties in a data sample have an inter-connected structure, and if we can capture the number of changes to convert one structure to another, then we can capture the differences between these samples. 
Based on this intuition, \system\ starts by constructing a \textit{hierarchical attributed relational graph}, called (\harg), with a common hierarchy for all data samples. The choice of graph as a representation was influenced by the need for 
% \begin{enumerate*}
%     \item graph being the best data structure to capture information from connected structures,
%     \item based on observations~\ref{itm:mmir-relationship} and \ref{itm:mmir-different-valued-zp}, 
    a data structure with representation-invariant encoding mechanisms that can capture the syntactic similarities between different values.
    % was necessary.
% \end{enumerate*}

\begin{definition}[Hierarchical Attributed Relational Graph] % (\harg)
\label{defn:harg}
\harg~ is a specific type of ARG in the form of a multi-level tree with $|h|$ levels. It consists of a root node, multiple levels of nodes and edges emanating from it, and specific type of leaf nodes. Nodes at level $h$ are denoted by $N^h$.
\end{definition}

\newcommand{\non}{N} %normalNodes
\paragraph{\textbf{CONSTRUCT-\harg}}
Each data sample is represented as \harg, following the steps:
\begin{enumerate}[leftmargin=*, topsep=0pt] % 
    \item The graph starts with a single node at level 0 $(h = 0)$ containing a common label across all data samples in the same application domain: $l(\non^0)=\{ROOT\}$. % \alt{term} (\textsc{content/ object/ root})
    \item Level 1 nodes constitute the object-properties of the data sample itself where the property name is the edge label, and the property value is the node label: 
    \[l(\non^0, \non^1) = \smallOP_p, l(\non^1) = z_{p}.\]
    With the exception of $o_p$ being an entity,  $\non^1$ would be a leaf node. For entities, we define the edge label as $l(\non^0, \non^1) = \{hasEntity\}$.
    \item In case a set of $\smallOP_p$ describes the attributes of an entity, $\non^k (k \geq 1)$ will be a pointer to 
    % another object
    the attribute properties of that entity, whereas $l(\non^k)$ = \{\emph{entity-type}\}. 
    \item We categorize entities in two groups for each data sample: \textit{primary}, and \textit{secondary}. 
    Level 1 of \harg~ only contains primary entities. 
    % (Definition~\ref{defn:primary-entity}).
    %
    \item Level 2 and subsequent levels contain the attribute values of the entities in the previous level with 
    \[l(\non^k, \non^{k+1}) = \smallOP_p (k \geq 1),  and\] \[l(\non^k) = z_p (k \geq 2).\]
    From Definition~\ref{defn:reln}, for \textattrname{relation} properties,  %$\smallOP_p = 
    $\langle R, S, Arg \rangle$ where entity-pointer $S$ is at level-$k$ and entity-pointer $Arg$ is at level-$(k+1)$, 
    \[l(\non^k, \non^{k+1})=R, l(\non^k)= {S}, l(\non^{k+1})={Arg}.\]
    \item There can be edges between entities in the same level with 
    % property-name as the edge label. 
    \textattrname{relation} properties, $R$.
    With nodes $\non^k$ and $\non^r$,\\ $l(\non^k, \non^r) = R$, where $l(\non^k) \neq l(\non^r)$ but $k=r$.
    \item The leaf nodes of \harg~ always contain a property-value or a NULL value for $z_p = \{\phi\}$.
\end{enumerate}

Figure 
% \ref{fig:harg}
\ref{fig:qg-and-cg}
demonstrates two examples of hierarchical attributed relational graphs from the \mmirdataexp dataset. $R1$ and $R2$ refer to two different data samples. For the leaf nodes $T2, M1,$ and $T3$ in $R2$, $z_p = \{\phi\}$. \textattrvalue{Wear}, and \textattrvalue{riding} refers to the \textattrname{Relation} property, where \textattrvalue{Persons} are subjects, and \textattrvalue{Clothes} and \textattrvalue{Motor-vehicles} are arguments. 

% \begin{definition}[Primary Entities]
% \label{defn:primary-entity}
% We consider entities that take the role of a subject in terms of a verb as primary entities:
% \begin{enumerate}
% \item for visual modalities, entities that control the action or relation properties,
% \item for textual modalities:
% \begin{enumerate*}[leftmargin=*, topsep=0pt]
%     \item for phase structure grammars, an immediate dependent of the root node \cite{rohrmeier2011towards},
%     \item for dependency grammars, an immediate dependent of the finite verb \cite{osborne2019dependency}, % TODO: re-check citation in paper
% \end{enumerate*}
% \item for database records, the entities from the tables with \textit{no foreign key constraints}. %, and the primary key element becomes the entity pointer.
% \end{enumerate}
% \noindent Secondary entities include any entities not satisfying the conditions of primary entities including \textit{objects, verb arguments, and themes}.
% \end{definition}

%
% https://www.thoughtco.com/object-in-grammar-1691445#:~:text=In%20English%20grammar%2C%20an%20object,the%20creation%20of%20complex%20sentences.
\begin{definition}[\textattrname{Relation} between Objects]
\label{defn:reln}
% Object-properties describing a relationship or action $R$ between two \cmt{objects/} entities $S$ (initiator) and $Arg$ (outcome/ receiver/ modifier) are defined as \textattrname{relation} properties, and the property-value is defined as a triplet of $\langle R, S, Arg \rangle$. 
\noindent For a $n$-ary relationship $R$, identifiers associate each action with multiple entity arguments, $Arg_1$, $Arg_2$, $\dots$, $Arg_i$, $\dots$, $Arg_n$ with \emph{role} $R^i_o$. 
% besides the subject $S$, and each of those arguments has a \emph{role}, $R^i_o$. 
$n$-ary relationships are broken into multiple binary relationships with 
\[l(\non^k, \non^{k+1})=\{R: R^i_o\}, l(\non^k)= {S}, l(\non^{k+1})=\{Arg_i\}.\] 
\end{definition}
%

%%%%%%%%%%%%%%%%%%%
% \noindent 
% \textit{Assumptions.} 
We made two assumptions for the generation process: %\harg~ generation process:
\begin{enumerate*}[label=(\textbf{\Roman*}), topsep=0pt, leftmargin=*]%, leftmargin=2 pt, label=(A\thesection.\arabic*)
\item we assume prior knowledge of the system-specific properties, 
% and that they have been extracted with appropriate property-identifiers, 
% (Section~\ref{sec:prop-identifiers}), 
\item the entity types for node labels are system-specific and must be consistent through the lifetime of the system. This assumption is valid since the property identifiers from each modality would be system-specific and extracted object types would be consistent across data samples.
\end{enumerate*}

\paragraph{Graph Embedding}
For calculating the graph embedding, first, Graph Convolutional Networks (GCN) \cite{kipf2016} are used on the \harg~ to obtain the node embeddings. GCN is representation-invariant and allows us to account for different kinds of labels for nodes and edges. %when ground truths are available. 
It is also inductive and allows computing the node embedding for any unseen graph following the GCN operation, which makes it a great choice for variable-sized \system\ graphs. Then, a global context-aware attention network is used to combine the node embeddings into a graph embedding. This allows \system\ to learn the importance of each feature in the similarity determination as part of the end-to-end network. 

%%%%%%%%%%%%%%%%%%%%%%%%%% Mid-section %%%%%%%%%%%%%%%%%%%%%%
\subsection{Similarity Label Generation with Content Edit Distance}
\label{subsec:weak-label-gen}

\newcommand{\propled}{\textsc{prDist}} %{Prop-List-Edit-Distance}
\newcommand{\hashcmp}{\textsc{hComp}} %{Hashmap-Compare}
\newcommand{\eplv}{\textsc{EPL(V)}}%{V_{EPL}}
\newcommand{\eple}{\textsc{EPL(E)}}
\newcommand{\epll}{\textsc{EPL(l)}}
\newcommand{\eplgraph}{g_\mathrm{epl}}

\system\ further defines a new distance metric, Content Edit Distance (\ced) using a variation of the Munkres' algorithm \cite{riesen2007bipartite} to calculate the amount \alt{number} of edits (changes) for optimal alignment of the query-example \harg~ to \harg~ of another data-sample. \ced~ is considered as weak label for the retrieval task for two reasons:
\begin{enumerate*}
    \item Munkres' algorithm is suboptimal as it only calculates approximate edit distance values,
    \item the quality of \harg~ rely on the choice of primary entity selection which can be noisy.
\end{enumerate*}
Our intuition was graph edit distance (GED) calculation algorithms (A*-search, VJ, or Beam) would be enough to calculate the number of changes after we have build the {\harg s}, but we made following observations.
\begin{enumerate}[label=\textbf{O\thesection.\arabic*}, topsep=0pt, leftmargin=*, resume=obsmmir] % just resume or 
\item \label{itm-rcost}
Different nodes and edges in \harg~ have different change cost. User should be allowed to specify individual property replacement cost. %which gives rcost+=,
\item \label{itm-ged-algo-speed-diff}
GED calculation algorithms differ in speed based on the number on nodes and \harg~ contains variable sized graphs. %which gives Munkers and NN,
\item \label{itm-dependency-betn-levels}
Object-properties such as, \textattrname{relation} have dependency between different levels of \harg~ and should not be considered individually during the change estimation. For example, for \textit{person wearing clothes}, edit cost for \textattrvalue{person} and \textattrvalue{cloth} should be considered together between different data-samples. 
% people and cloth entity should not be considered separately % which gives cumulative-munkers.
%
\item \label{itm-list-edit-dist}
Considering \ref{itm:mmir-single-multi-value-zp}, we cannot calculate the edit cost of certain properties just by replacing or deleting them since they have multiple number of values in their value-set. % which gives LED
\end{enumerate}
%

%%%%
% Intuitively, we can just compare the property-values between two data samples and can sum the (number of mismatch $*$ replacement cost for each property) to calculate the amount of edits, following \ref{itm-rcost}. But due to the presence of list-properties such as \ref{itm-list-edit-dist}, we needed to include a method for \textbf{list-property comparison}. There are two cases for such exceptions where the user is interested in: (1) an ordered comparison of the list values, or (2) and unordered comparison\cmt{is sufficient}.\\
% %
% \textbf{\propled.} For an \textit{ordered comparison of a value-list} we can consider the problem as the \textit{Levenshtein distance} problem where the distance between two lists is the minimum number of \textit{single-property edits} (i.e., insertions, deletions, or substitutions) required to change one list into the other. Each of these operations has a unique cost.\\
% %
% \textbf{\hashcmp.} For an \textit{unordered comparison}, we can use \textit{hash table} to iterate through the lists and sum the remaining (entries $*$ corresponding property replacement-cost) in the hash table to calculate the total amount of edit to replace the property.\\
For properties with list values, we consider two types of comparison: \textbf{(\propled)} ordered comparison with modified \textit{Levenshtein distance}, and \textbf{(\hashcmp)} unordered comparison with \textit{hash table}.
%%
% For solving \ref{itm-rcost} and \ref{itm-list-edit-dist}, 
Summing the cost of edits for all the properties between two data-samples ignores the inter-connected structure among the properties. 
In Figure~\ref{fig:qg-and-cg}, the graph from $R2$ has two persons, and while comparing with $R1$ we would want to know the minimal edit cost by considering which person in $R2$ is closer to the person described in $R1$. % !!!! TODO: Double check what I said here.
\textsc{Content Edit Distance} calculates the cost for the minimal cost alignment of one data-sample to another.
Since only property values in leaf nodes in a \harg~ have direct replacement cost, 
% For calculation of \ced, 
we propose a new kind of vertex in \harg, \textit{Entity-with-Property-in-Leaf (EPL) vertex} (Definition \ref{defn:eplv}) for calculating the cost for an individual object assignment. 
% Q1. Why are the entities as child are independent? - as EPL-vertex gives the replacement cost of that entity.
Given
$\eplv$ is the finite set of EPL Vertices, 
$\eple \subseteq \eplv \times \eplv$ is the set of edges, and $\epll \subset l$ is 
% the same labeling function we used for an ARG, 
the labeling function,
a \harg~ is now defined as:
\[
\eplgraph = (\eplv, \eple, \epll)\]
% this basically means now the edges are defined between epl vertices.  
%
\begin{definition}[Entity-with-Property-in-Leaf Vertices]
\label{defn:eplv}
A node labeled with object-type ($A$) with their outgoing edges
% $e \subset \eple$
labeled with object-properties ($\smallOP_p$) and the connected leaf nodes labeled with property-values ($z_p$) are considered as \textsc{Entity-with-Property-in-Leaf} (\textbf{EPL}) Vertex, $\eplv$.
% An EPL vertex can be connected to other nodes representing a pointer to another EPL vertex or an entity.
A node without any leaf nodes is also considered as an EPL vertex. An EPL vertex can be connected to other EPL vertices and have its own cost functions. \cbd{Two cost functions are known/ pre-defined for all properties in an EPL vertex:
\begin{enumerate}
    \item $rcost$($\smallOP_p$) is the replacement cost of property-value of $\smallOP_p$.% in $d_c$ to property-value of $\smallOP_p$ in $d_q$.
    \item $icost$($\smallOP_p$) is the insertion cost for missing value of $\smallOP_p$.% in $d_c$.
\end{enumerate}
}
\end{definition}
% 

% \noindent
% Plain-Munkres and Adjacency-Munkres for \harg~ Matching.}
\paragraph{\textbf{Munkres Algorithm for CED calculation}}
% \harg~ Matching.}
%%%%%%
We consider the CED calculation as an assignment problem and adopted the bipartite graph matching method in \cite{riesen2007bipartite}. Compared to the exponential time-complexity of A*-search, Munkres' \cite{riesen2007bipartite} algorithm has a polynomial time complexity. Estimating content edit distance instead of a simple property-to-property comparison allows the flexibility to consider the dependency between properties and graph levels. %\cmt{and different levels of \harg}
%%%%%%%
Given the non-empty \har~ graph from query-example, $\eplgraph^q$ = $(\eplv^q, \eple^q, \epll^q)$ and the \har~ graph from the compared data-sample, $\eplgraph^c$ = $(\eplv^c, \eple^c, \epll^c)$,
where\\ $\eplv^q = \{u_1, \ldots, u_n\}, \eplv^c = \{v_1, \ldots ,v_m\}$, the Munkres' algorithm would output \ced~($\eplgraph^q$, $\eplgraph^c$). % u_i and v_j both are some z_p, but to show different node values we denote with u_i, v_j
We made the following adjustments to the Munkres' algorithm in \cite{riesen2007bipartite}.
\begin{enumerate}[left=0pt]
    \item EPL-vertices in the query graph need to be aligned to the data-samples, hence we will fix the assignment size $k$ to $|\eplv^q|$. \cbd{Initialize $k$ = $n$.}
    %%%%%%
    \item For data retrieval, the entities and relations in query graph \cbd{($\eplgraph^q$)} needs to be in comparison-graph\cbd{ ($\eplgraph^c$)}, otherwise indicates missing property. So there is no need to add dummy nodes to $\eplgraph^q$. 
    Formally, if $n > m$,  % or m>n 
    only the costs for 
    % $max\{0, n-m\}$ node deletions and 
    $max\{0, m-n\}$ node insertions 
    % $c(\epsilon \rightarrow v_j)$ \alt{icost} \alt{(to $g^c_v$)} 
    have to be added to the minimum-cost node assignment.
    % So there is no need to add dummy nodes.
    %%%%%%
    \item Next, the $n \times m$ cost-matrix $C$ is generated.
    (1)     For different type of \cmt{properties, or } objects $A$ in $u_i$ and $v_j$ the replacement cost is set to $\infty$.
    (2) The cost for a single object assignment $C_{i,j}$ is calculated by comparing the property values $z_p$
    (normal-comparison and list-comparison)
    % methods mentioned above
    in EPL-vertex $u_i$ and $v_j$.
    % (optimal assignment of an object in query-example $u_i$ to an object in a data-sample $v_j$)
    % both u_i and v_j belongs to same property and their values are from same property-value set z_p, can be same or different. %%% can't be based on incoming edge on them as edge can have wear or is wearing, and wont match.
    %% TODO: want to give a reason/ maybe intution.
    %%% OMG! I almost got into the drain again - thinking how would i know what is the replacement cost of R1 or P1, i thought i only knew last level replcement cost. BUT NO! We know cost of each levels nodes cost based on only the feature nodes. So the cost I wrote on lab board was based on only the feature node replacement, not the other entities as children.
    \item To accommodate for \ref{itm:mmir-different-valued-zp}, while applying Adjacency-Munkres, we set the default cost of an edge replacement $c(e_{u_i} \rightarrow e_{v_j})$ based on the Wu-Palmer distance between Synsets of $l(e_{u_i})$ and $l(e_{v_j})$. $e_{u_i}$ denotes all edges connected to $u_i$ and $e_{v_j}$ denotes all edges connected to $v_j$. In general, any language embedding can be used instead of Synsets. % Verb synsets of the edge-label tokens. % if >0.9 cost = 0 etc etc 
    \begin{gather}
    \label{eqn:edge-replace-cost}
        c(e_{u_i} \rightarrow e_{v_j}) = 1/{wpdist(s_{l(e_{u_i})},~ s_{l(e_{v_j})})} %\\
        % C_{i,j} = C_{i,j} + min (\sum  c(e_{u_i} \rightarrow e_{v_j}))
    \end{gather}
\end{enumerate}
%
% \noindent
% Plain-Munkres and Adjacency-Munkres for \harg~ Matching.}
\paragraph{\textbf{Cumulative-Munkres}}
Using Adjacency-Munkres from \cite{riesen2007bipartite} allows us to find the optimal assignment of each EPL vertex without taking into account the dependency among them \ref{itm-dependency-betn-levels}. We utilize the levels from \harg~ to include the dependency information into the cost-matrix.
So for every $C_{i,j}$ in the cost matrix from adjacency-munkres denoting an assignment of $u_i$ to $v_j$, we add their parent EPL-vertices assignment cost to $C_{i,j}$, starting from EPL-vertices in level-1. In the remainder of
this paper, we will call this method \textsc{Cumulative-Munkres} since it uses the cumulative cost of the parent and child nodes to preserve the dependency information.
% \[C_{i,j} = C_{i,j} + C_{parent(i),~ parent(j)}\]
\cbd{
For example, in Figure~\ref{fig:qg-and-cg}, we have following costs, $C_{4,3} = 1$, 
%(assigning $u_4$ to $v_3$) 
$C_{4,7} = 3, C_{2, 2} = 6,$ and $C_{2, 6} = 3$, with $parent(u_4) = u_2, parent(v_3) = v_2$, and  $parent(v_7) = v_6$. So, the new cost would be, $C_{4,3} = 7, C_{4,7} = 6$.
% \[C_{4,3} = C_{4,3} + C_{2,2} = 7\] 
% \[C_{4,7} = C_{4,7} + C_{2,6} = 6\]
Although individually $v_3$ is a better assignment for $u_4$ than $v_7$, when we take the parent cost into account, $v_7$ is a cheaper assignment for $u_4$.
}

\subsection{Approximate CED Inference}
Finally, we propose to use an end-to-end neural network model, SimGNN \cite{bai2018graph} to learn an embedding function to map $d_q$ and $d_c$ into a similarity score based on the \ced~ score. User requirements (such as relationships between properties, searching in a time range, or within a specified location, etc.) and different system constraints are considered as function parameters with appropriate replacement costs while calculating \ced. 
%%%%
Similarity scores for training the model are derived by normalizing the distance scores \cite{qureshi2007graph} and applying an exponential function on the normalized score. (Line \ref{algo:line:ced-25-calc-sim} in Algorithm~\ref{algo:ced}).
%%%
The embedding function outputs a number of interaction scores between the pair of graphs using Neural Tensor Networks (NTN) \cite{NIPS2013_5028} on the graph embeddings. 
% In addition, SimGNN augments the graph level interaction score with local information by calculating histogram features from a pairwise node interaction score between the node embeddings.
Finally, a multi-layer fully connected network is applied to 
% gradually decrease the number of similarity scores into a singular one, 
learn a single similarity score from the interaction scores, which
%, $\hat{s_{ij}} \in \mathbb{R}$. The predicted score 
is compared against the weak \ced\ labels or the ground-truths using mean squared error loss. %the following mean squared error loss function:
\begin{equation}
    \mathcal{L}_{mse} = \frac{1}{|D|}\sum_{d_q, d_c\in D}{(\hat{s} - s(d_q, d_c))^2}
\end{equation}
where D is the set of data samples from the repository or the stream, $\hat{s}$ is the predicted similarity score, and $s(d_q, d_c)$ is the
ground-truth similarity between $d_q$ and $d_c$. This similarity score allows us to rank the data samples against the query example.
%%%%%%%%%%%%%%%%%%%%%% Mid-section %%%%%%%%%%%%%%%%%%%%%%%%%
\subsection{\system\ algorithm}
% https://www.geeksforgeeks.org/check-two-unsorted-array-duplicates-allowed-elements/

%%%
\begin{algorithm}[h]
% \SetAlgoLined
% \DontPrintSemicolon
%
\begin{algorithmic}[1]
\Require{Query example and a single Data sample, $d_q$ and $d_c$ \newline
Replacement cost for property $\smallOP_p$, \rcost \newline 
Insertion cost for property $\smallOP_p$, \icost}
%%%%%%%
\Ensure Similarity score between $d_q$ and $d_c$, SIM ($d_q, d_c$)
% \KwOut{Content Edit Distance, \cedoutsidealgo}%~ between $d_q$ and $d_c$}
%
%

%
% \ced $\gets$ 0\;
%%%Comment%%%% \tcc*[r]{$\bigOP^j$ is set of object-properties in data sample $d_j$}
% $\{o^q_p, z^q_p\} 
%%%%%%%%%%%%%%%%%% line %%%%%%%%%%
\State $ \bigOP^q \gets \textsc{Prop}(d_q)$,  
% \tcc*[r]{in prob-def we introduced a func. Prop(d)}
% $\{o^c_p, z^c_p\} 
$\bigOP^c \gets \textsc{Prop}(d_c)$                      \label{algo:line:ced-1-prop-id}\;
%%%%%%%%%%%%%%%%%% line %%%%%%%%%%
\State $g^q \gets \textsc{Construct-HARG}~ (\bigOP^q)$\; %(o^q_p, z^q_p)$\;
                                                \label{algo:line:ced-2-cons-harg-query}
\State $g^c \gets \textsc{Construct-HARG}~ (\bigOP^c)$\; %(o^c_p, z^c_p)$\;
                                                \label{algo:line:ced-3-cons-harg-compare}
\If{training}                                  \label{algo:line:ced-4-if-training}
\State $\eplgraph^q,~ \eplgraph^c \gets \textsc{Discover-EPLV} (g^q,~ g^c)$\;
                                                \label{algo:line:ced-5-gepl-create}
% COMPUTE-COST-MATRIX-START\;
\State $\CostMat \gets \phi$ \;                        \label{algo:line:ced-6-ced-initialize}
\ForEach{$u_i \in \eplv^q$}                    \label{algo:line:ced-7-for-ui}
\ForEach{$v_j \in \eplv^c$}                    \label{algo:line:ced-8-for-vj}
    \If{\type ($u_i$) $\neq$ \type ($v_j$)}  
                                                \label{algo:line:ced-9-not-same-type-object}
        \eplvcost = $\infty $
    \EndIf
    \ForEach{$\smallOP_p \in u_i$}             \label{algo:line:ced-10-for-each-op-in-ui}
        %%%Comment%%%% \tcc*[r]{property is absent in compared data-sample, need to be inserted. Higher cost than replacement}
        \If{$\smallOP_p \notin v_j$}          \label{algo:line:ced-11-missing-property}
            \eplvcost~ += \icost
        % \EndIf
        %%%Comment%%%% \tcc*[r]{compare the property-values}   
        \ElsIf{\type($z_p$) is not list}      \label{algo:line:ced-12-prop-is-list}
            % \tcc*[r]
            \mynewlinecomment{\zpinq is value of $\smallOP_p$ in vertex $u_i$}
            \If{\zpinq $\neq$ \zpinc}          \label{algo:line:ced-13-not-equal-start}
                \State \eplvcost += \rcost 
            % \EndIf                                    
            \label{algo:line:ced-14-not-equal-end}
            \Else
                \State \eplvcost += 0               \label{algo:line:ced-15-equal-prop-value}
            \EndIf
        
        \Else           \label{algo:line:ced-16-list-prop}
            % \State 
            \begin{align*}
                &\eplvcost~ +=\\
                \{\ordcmp &* \listPropED{\zpinq, \zpinc} +\\
                (1 - \ordcmp) &* \listPropComp{\zpinq, \zpinc} \}
            \end{align*}
            % \eplvcost~ +=~  \\ \{\ordcmp$*$\listPropED{\zpinq, \zpinc} +\\ 
            % (1 - \ordcmp)$*$\listPropComp{\zpinq, \zpinc} \}
                                                \label{algo:line:ced-17-list-prop-calc}
        \EndIf
    \EndFor
    % \tcc*[r]{The further the edge labels are in meaning, the lower is the distance score, and the higher the edit cost} % TODO: if wpdist gives higher score for being further then change it to just val from 1/val
    % \tcc*[r]{calculate for all connected edges}
    % $c(e_{u_i} \rightarrow e_{v_j}$) = $\frac{1}{wpdist(s_{l(e_{u_i})},~ s_{l(e_{v_j})})}$ 

    \State \eplvcost = \eplvcost + $ min \{\sum  c(e_{u_i} \rightarrow e_{v_j})$\} \;
                                                \label{algo:line:ced-18-add-edge-cost}
    %%%%% COMPUTE-COST-MATRIX-END\;
    \EndFor
\EndFor
%
% Apply Munkers' algorithm \;
% Calc. \ced \;
% \ced = \munkersType $*$ \textsc{Adjacency-Munkres} +\\ (1 - \munkersType) $*$ \textsc{Cumulative-Mukres}
\If{\munkersType}                              \label{algo:line:ced-19-cum-munkers-start}
    \ForEach{$u_i \in \eplv^q$}                \label{algo:line:ced-20}
        \ForEach{$v_j \in \eplv^c$}            \label{algo:line:ced-21}
                \State $u_{\hat{i}} = parent(u_i), \hfill 
                v_{\hat{j}} = parent(v_j)$ \;   \label{algo:line:ced-22-parent-calc}
            \State $C_{i,j} = C_{i,j} + C_{\hat{i}, \hat{j}}$
                                                \label{algo:line:ced-23-add-parent-cost}
        \EndFor
    \EndFor
\EndIf
\State \ced (~$g^q, g^c$~) = \textsc{Munkres} (\CostMat)\;
                                                \label{algo:line:ced-24-apply-munkres}
\State n\ced = $\frac{\ced (~g^q, g^c~)}{(|g^q|+|g^c|)/2}$ 
\State SIM ($d_q, d_c$) = $e^{-n\ced}$ \;              \label{algo:line:ced-25-calc-sim}
% Add \{($g^q, g^c$), SIM ($d_q, d_c$)\} to training data\;
%
% \EndIf
\Else                   \label{algo:line:ced-26-testing-phase}
    \State SIM ($d_q, d_c$) = \simgnn (~$g^q, g^c$~) \;                            
                      \label{algo:line:ced-27-apply-simgnn}
\EndIf
%
% \Return SIM ($d_q, d_c$)
\end{algorithmic}
% \caption{Content Edit Distance (\cedoutsidealgo)}
\caption{\system}
\label{algo:ced}
\end{algorithm}
%
%%%%
Algorithm \ref{algo:ced} presents the pseudocode of our retrieval algorithm \system\, which takes two data samples as input and returns the similarity score between them as output.\\
\textbf{Line \ref{algo:line:ced-1-prop-id}} Extract the set of properties and their values, $\bigOP^j$ from data-sample $d_j$ using the \cmt{pre-defined} modality-specific property-identifiers.\\
\textbf{Lines \ref{algo:line:ced-2-cons-harg-query} - \ref{algo:line:ced-3-cons-harg-compare}} Construct the Hierarchical Attributed Relational Graphs using the identified properties following the steps in Section~\ref{subsec:data-ingestion-w-graphs}.\\
\textbf{Lines \ref{algo:line:ced-4-if-training} - \ref{algo:line:ced-25-calc-sim}} During training, generate the \ced\ as weak label using the Munkres algorithm. \ced~ is used to calculate the similarity score, and this pair of data-samples and the similarity score is added as training sample for SIMGNN.\\
\textbf{Line \ref{algo:line:ced-5-gepl-create}} Calculate the EPL-vertices in the {\harg s}, $\eplgraph$.\\
\textbf{Line \ref{algo:line:ced-6-ced-initialize}} Initialize an empty $n \times m$ cost-matrix \CostMat.\\
\textbf{Lines \ref{algo:line:ced-7-for-ui} - \ref{algo:line:ced-8-for-vj}} Iterate through all the vertices in $\eplv^q$ and $\eplv^c$ and compare the properties in each vertex to assign the costs.\\
\textbf{Line \ref{algo:line:ced-9-not-same-type-object}} For different types of object, set the cost to $\infty$, not allowing different types of object to be aligned. \\
\textbf{Line \ref{algo:line:ced-11-missing-property}} If a property in $u_i$ is absent in $v_j$, it needs to be inserted in $v_j$. Increment the cost-matrix value by the insertion-cost.\\
% It should have higher cost than just replacement cost.
%
\textbf{Lines \ref{algo:line:ced-12-prop-is-list} - \ref{algo:line:ced-15-equal-prop-value}} If the property is not a list, then just compare the values in $u_i$ and $v_j$. If they mismatch, add the replacement cost to the cost-matrix, otherwise nothing is added.\\
\textbf{Line %s \ref{algo:line:ced-16-list-prop} - 
\ref{algo:line:ced-17-list-prop-calc}} If the property is a list, we need to compare them either with a Levenshtein distance (ordered comparison) or with a hashmap (unordered comparison) from Section~\ref{subsec:weak-label-gen}. \ordcmp\ is a control variable to specify what kind of comparison is required. The overall cost is added to cost-matrix.\\
\textbf{Line \ref{algo:line:ced-18-add-edge-cost}} For applying Adjacency-Munkres, the minimum edge replacement cost is added to the cost matrix using Equation~\ref{eqn:edge-replace-cost}.\\
\textbf{Lines \ref{algo:line:ced-19-cum-munkers-start} - \ref{algo:line:ced-23-add-parent-cost}} If Cumulative-Munkres is required (set by \munkersType), cost-matrix entry of the parent vertices are added to each $C_{i,j}$.\\
\textbf{Line \ref{algo:line:ced-24-apply-munkres}} Apply the Munkres algorithm to calculate the optimal assignment based on \CostMat, and the associated cost is the \ced.\\
\textbf{Line \ref{algo:line:ced-25-calc-sim}} Normalize \ced\ to the graph sizes and apply an exponential function to convert it to a similarity score in the range of (0, 1]. Add it to training sample for SIMGNN.\\
\textbf{Lines \ref{algo:line:ced-26-testing-phase} - \ref{algo:line:ced-27-apply-simgnn}} During inference, apply the learned mapping function to predict the similarity score from the {\harg s} and rank based on that.\\
\noindent \textbf{Generalization:} 
\begin{enumerate}
    \item Algorithm~\ref{algo:ced} assumes that the edge labels for level 0 are fixed to $hasEntity$ and $metadata$ with granularity (such as $time, location$, etc.). These are flexible and can be set to any labels in \system\ as long as it is consistent throughout the lifetime of the system.
    \item Object-types are assumed to be system-specific, and can be variable across different systems and applications. \system\ can handle any labels for entity-type since the retrieval result does not depend on it. The comparison between properties is affected by it which remains valid as long as the same heuristics are maintained for all modalities in a system.
    \item \system\ is capable of handling different replacement costs and insertion costs for properties in different application domains. 
    \item For the edge replacement cost, any language embedding will work as long as the objective function places semantically similar tokens closer to each other.
\end{enumerate}

%%%%%%%%%%%%%%%%%%%%%%%%%%%%%%%%%%%%%%%%%%%%%%%%%%%%%%%%%%%%%%%%%%%%%%%%%%%%%%%%%%%

\section{Human Attribute Recognition from Unstructured Text (HART)} 
\label{sec:text-feature-extraction}
We now describe the property identification technique for unstructured texts to extract \textit{attribute-based properties} \cmt{human properties} from large text documents. Our algorithm considers the full document as input and reports a \textit{collection} of object-properties and their set of values, 
%$\langle\langle o_p, z_p \rangle\rangle$ 
as output. To this end, we first identify the candidate sentences $C_s$ from a collection of sentences $T_s$ by searching for the key-phrases ($\kp$)
using pre-trained language representation models and lexical knowledge bases. 
Then, we propose individual property-focused models to extract the attributes and their corresponding values using the syntactic characteristics (i.e., parts-of-speech) and lexical meanings of the tokens in the \textit{Candidate Sentences}. Our heuristic search algorithm, \textsc{\posalgo} iteratively checks the tokens in the candidate sentences and based on the assigned tags in accordance with their syntactic functions \cmt{in English language} \alt{linguistic structure in English} identifies the properties in $\mathcal{O}_H$ and their values.

\subsection{\textbf{Candidate Sentence Extraction}}
\label{subsec:cand-sent-ex}
A naive approach to this task would be to consider it as a 
% binary classification task 
supervised classification problem % or learning problem
given enough training data. Since during this work, the primary goal was to define on-demand models that works in absence of training data,
% our primary objective was to define models that require minimal amount of training data and have on-demand response rate, we build our extraction models on top of pre-trained language representation models and lexical knowledge bases while searching for linguistic attribute indicators.
we designed this as a similarity search problem using pre-trained and lexical features, where the similarity between sentence and key-phrase needs to reach an empirical threshold.
% Similarity can be defined in different ways:
% \begin{enumerate*}
%     \item pattern matching,
%     \item semantic similarity, or
%     \item syntactic similarity.
% \end{enumerate*}
We now proceed to describe the different methods \alt{similarity metrics} used to identify $C_s$.
%
% We experimented with several different combinations of the above-mentioned alternatives, and compared them by building a multi-label classifier for \textit{clothes} attribute value prediction over the document. The results of these experiments are summarized in Table ~\ref{tab:clothes-various-results}. % Due to space constraints, we provide a brief overview of these alternatives, and point to the full description in the relevant papers.
%
%
%
% \setlength{\parindent}{0pt}
\paragraph{\textbf{Pattern Matching}}
\label{par:text-pattern-match} 
As a baseline heuristic model, we implemented the \textbf{\textsc{Regular Expression (RE)}} Search on $T_s$. 
% We define the regular expressions based on a manual observation \alt{analysis} of the different styles around the key-phrases in $\setkp$~ from a limited amount of un-annotated data.
Since we consider all sentences in the document as input corpus, if it describes multiple persons, this model captures all of the sentences describing a person as $C_s$. 
% without any discrimination. 
% We identify the individual mentions in property identification stage.  % rethink these 2 lines
Individual mentions are differentiated in later stages.
For RE, $\simhart(\kp, s) \in \{0, 1\}$. Given the key-phrase $q_H$, the RE pattern searches for any sentence mentioning it:

% \hrule 
% \vspace{1 pt}
%%% anything+$q_h$+anything except . (one or more times).
%%%% tokens around $q_H$: \hfill
\begin{center}
\verb|[^]*|$q_H$\verb|[^.]+|
\end{center}
% \verb|[^.]*wear[^.]*|
% \vspace{1 pt} 
% \hrule
%

\paragraph{\textbf{Similarity using Tokens}}
Similarity between $\kp$ and $s$ is calculated based on the similarities between tokens $w \in s$ and $\kp$. 
% For each token $w in s$ and for each phrase $\kp \in \setkp$, we use the same token representation model. 
A single model is used to embed both $w$ and $\kp$ into the same space.
We used two different token representation models for token to query phrase similarity.
\begin{equation}
\label{equ:cs-sim-using-tokens}
    \simhart (\kp, s) = \max_{w \in s} \simhart(\kp, w)
\end{equation}

(a) \emph{\textbf{Word Embedding.}} % For each token $w in s$ and for each phrase $\kp \in \setkp$, we use the same token representation model.
    Tokens in each sentence and in the key-phrase are represented by \textbf{\textsc{Word2Vec}} \cite{mikolov2013efficient} embeddings. 
    % word embeddings from language representation models.
    If there are multiple tokens in a key-phrase, the average of the embeddings are used. We use cosine similarity as the distance metric.
    Given $u_{q_H}$ and $u_w$ are the final embedding vectors for $q$ and $w$,
    \begin{equation}
        \simhart(\kp, w) = cos(u_{\kp}, u_w) = \frac{u_{\kp} \cdot u_w}{\lVert{u_{\kp}}\rVert \cdot \lVert{u_w}\rVert}
    \end{equation}
    
%
%
%
% WordNet is a lexical database of semantic relations between words in more than 200 languages. WordNet links words into semantic relations including synonyms, hyponyms, and meronyms. The synonyms are grouped into synsets with short definitions and usage examples.
(b) \emph{\textbf{Word Synsets.}}
Tokens and key-phrases are represented by \textbf{\textsc{Wordnet}} \cite{wordnet98} synsets in \texttt{NOUN} form. For similarity/distance metric, we used the Wu-Palmer similarity \cite{wu1994verb}. 
% It calculates the similarity based on how similar the word senses are and where the Synsets occur relative to each other in the hypernym tree. 
Given the synsets of $q$ and $w$ are $s_{\kp}$ and $s_w$, %the similarity score is,
\begin{equation}
    \simhart(\kp, w) = wpdist(s_{\kp}, s_w) %= 2 * \frac{depth(lcs(s_{\kp}, s_w))}{depth(s_{\kp}) + depth(s_w)}
\end{equation}

\paragraph{\textbf{Classification Model}}
The similarity search problem is re-designed as a classification problem where the sentences are considered as input sequences, and the key-phrases are considered as labels. The probability of sequence $s$ belonging to a class $\kp$ is then considered as the similarity between a sentence and a key-phrase. To that end, following Yin et al. \cite{yin2019benchmarking}, we used pre-trained natural language inference (NLI) models as a ready-made zero-shot sequence classifier. The input sequences are considered as the NLI premise and a hypothesis is constructed from each key-phrase. For example, if a key-phrase is \texttt{clothes}, we construct a hypothesis \textit{"This text is about clothes"}.  
The probabilities for \textit{entailment} and \textit{contradiction} are then converted to class label probabilities.
%  we throw away "neutral" (dim 1) and take the probability of # "entailment" (2) as the probability of the label being true.
Then, both the sequence and the hypothesis containing the class label are encoded using a sentence level encoder Sentence-BERT \cite{reimers2019sentencebert} (\textbf{\textsc{SBert}}). Finally, we use the NLI model to calculate the 
% \cmt{posterior-not posterior, as now calculating class=entailment basically, that is likelihood, not posterior} 
probability $P$. Given SBERT embedding of a sequence $s$ is denoted with $B_s$,
\begin{equation}
    \simhart(\kp, s) = P({s ~\text{is~ about}~ \kp} ~|~ B_s, B_{\kp})
\end{equation}
%
% SBERT is fine-tuned on the combination of SNLI\cite{bowman-etal-2015-large} and Multi-Genre NLI\cite{williams-etal-2018-broad} dataset. 
% a modification of the pretrained BERT network that use siamese and triplet network structures to derive semantically meaningful sentence embeddings that can be compared using cosine-similarity. T

\paragraph{\textbf{Stacked Models}} % Candidate sentences are identified based on the combination of two approaches. 
While \textbf{RE} search relies on specific patterns and returns exact matches, the other models calculate a soft similarity, $0 \leq \simhart(\kp, s) \leq 1$.
Hence if initial results from \textbf{RE} search returns no result for all the key-phrases % clothes, %for the token \textit{wear[/s/ing]}, 
we %tries to identify words similar to \texttt{cloths} based on the 
use %Word Synset Comparison and the SBERT Classification model 
\textbf{\textsc{Wordnet}} or \textbf{\textsc{SBert}} model to identify semantically similar sentences to the key-phrases.
\subsection{\textbf{Iterative Search for Properties}}
\label{subsec:iterative-search-prop}
We now formally describe the \posalgo~ algorithm, which uses the models described in Section~\ref{subsec:cand-sent-ex} in the first stage. We start with the observations that led to the \posalgo~ algorithm. 

%
%%%%%%%%%%%%%%%%%%%%%%%%%% mid-mid-section %%%%%%%%%%%%%%%%%%%%%%%%%
\paragraph{Observations} We proposed \posalgo~ based on the following observations: 
\begin{enumerate}[label=\textbf{O\thesection.\arabic*}, topsep=0pt, leftmargin=*] % , leftmargin=*, topsep=0pt
    \item Common to \ref{itm:mmir-single-multi-value-zp}, 
    % some properties in $z_p$ have single value i.e., \textattrname{gender, race, height}, while other properties have a variable number of values i.e., \textattrname{clothes}.
    object-properties have single and multiple value contrasts.
    \item Some properties follow specific patterns such as \textattrname{gender} = \textattrvalue{\{male, female, man, woman, binary, non-binary, \ldots \}}, whereas some properties have variable values, as shown in \ref{itm:mmir-different-valued-zp}.
    \item Adjectives \textsc{(ADJ)} are used for naming 
    or describing characteristics of 
    a property, or used with a \textsc{NOUN} phrase to modify and describe it.
    \item Property values can span multiple tokens, but they tend to be consecutive.
    \item Property values for \textattrname{clothes} generally include the color, a range of colors, or a description of the material. 
    \item \textattrname{Clothes} usually is described after consecutive tokens with VERB tags, $V_{DG}$, such as, gerund or present participle \textsc{(VBG)}, past tense \textsc{(VBD)} etc.
    % , present tense \& third person singular \textsc{(VBZ)} etc.). % after \textstring{wearing}.
    If proper syntax is followed, 
    % a past history of a human profile 
    an entity is described with a \textsc{VBD} followed by a \textsc{VBG}. In most cases, mentioning \textstring{wearing}.  
    \label{obs:first_verbs}
    \item After a token with \textsc{VBG} tag, until any \textsc{ADJ} or \textsc{NOUN} tag is encountered, any tokens describing the set $P_{DCP}$, \{Determiner\cbd{(DET)}, Conjunction\cbd{(CONJ)}, Preposition\cbd{(PREP)}\}, or a Participle\cbd{(PTCP)}, or Adverb \cbd{(ADV)} is part of the property-name. An exception would be any \{participle, adverb, or verb\}, $P_{PAV}$  preceded by any \{pronouns \cbd{(PRP)} or non-tagged tokens\}, $P_{P\epsilon}$, which ends the mention of a property-name.
    \label{obs:property-name-decribe}
    % \item \textattrname{Clothes} are always mentioned at the end of the candidate sentence after describing other properties.
\end{enumerate}

    \algnewcommand\cands[1]{\textsc{extract-$Cs_{RE}$(#1)}}
    \algnewcommand\candsModel[1]{\textsc{extract-$Cs_{model}$(#1)}}
    \algnewcommand\TOK[1]{\textsc{Tokenize-Word(#1)}}
    \algnewcommand\POS[1]{\textsc{POS(#1)}}
    \algnewcommand\syn[1]{\textsc{synsets(#1)}}
    \algnewcommand\fndmtch[1]{\textsc{matchColor(#1)}}
    \algnewcommand\popopn[1]{\textsc{propName(#1)}}
    \algnewcommand\reForFiniteValuedProp[1]{\textsc{\repropvalues{#1}}}%{RE-PROP-VALUES} 
    \algnewcommand\reForVariableValuedProp[1]{\textsc{RE-Variable-valued-Props{#1}}} % returns partial sentence
    %%%
    \algnewcommand\CONCAT{\textsc{concat}}
    \algnewcommand\APPEND{\textsc{append}}
    \algnewcommand\LENGTH{\textsc{length}}
    \algnewcommand\search{\textsc{search}}
    %
    % Set Names
    \algnewcommand\SetOne{$P_{DCP}$}
    \algnewcommand\SetTwo{$P_{PAV}$}
    \algnewcommand\SetThree{$P_{P\epsilon}$}
    \algnewcommand\SetFour{$V_{DG}$}
    \algnewcommand\tempzp{$L_z$}%{${Temp}_{z_p}$}
    \algnewcommand\finiteValuedProp{$fo_p$}
    % keywords
    \algnewcommand\Continue{\textbf{continue}}
    \algnewcommand\Break{\textbf{break}}
    \algnewcommand\result{$\bigOP_H$} % subset of O
    \algnewcommand\tokens{$T_o$}   
    \algnewcommand\tags{$T_a$}
    \algnewcommand\index{i}
    \algnewcommand\word{$w_i$} 
    \algnewcommand\wordWithoutIdx{$w$} 
    \algnewcommand\postag{$t$}
    \algnewcommand\nid{$N_{idx}$}%{names\_index} o_p
    \algnewcommand\names{$N$}
    \algnewcommand\desc{$D$}%{descriptions}    z_p
    \algnewcommand\descColor{$d_{color}$}  % if description is a color
    \algnewcommand\wm{${S}_w$}%{word\_synset}
    \algnewcommand\colour{\emph{$COLOR_{syn}$}}
    \algnewcommand\mn{$s_{syn}$}
    \algnewcommand\PRP{\emph{PRP}}
    \algnewcommand\vrb{\emph{VERB}}
    \algnewcommand\Noun{\emph{NOUN}}
    \algnewcommand\none{\emph{NONE}}
    \algnewcommand\adj{\emph{ADJ}}

\begin{algorithm}[!htb] %\footnotesize
\begin{algorithmic}[1]
\Require{Collection of Sentences, $T_s$}
\Ensure{Collection of $\langle name, values \rangle$ pairs, $\langle\langle \smallOP_p, z_p \rangle\rangle$}, \result
\State \finiteValuedProp $\gets \{ \textattrname{gender, race, height}\}$ \;
\State \colour $\gets$ \syn{``color'', \Noun}[0]\;
\State $C_s \gets$ \cands {$T_s~, Q_H$} \;  \label{algo:line:cand-start}
\If{$C_s$ is $\phi$}
    \State $C_s \gets$ \candsModel {$T_s~, Q_H$}   \label{algo:line:cand-end}
\EndIf
\State \result $\longleftarrow \emptyset$
\mysamelinecomment{Collection of $\langle o_p, z_p \rangle \equiv \langle name, values \rangle$ pairs}
\ForEach{$s$ in $C_s$}                      \label{algo:line:posid-8}
    \ForEach{$o$ in \finiteValuedProp}      \label{algo:line:regex-for-value-start}
        % \tcc*[r]{\tempzp is last token in $s$ which is a property-value}
        \State \tempzp = \reForFiniteValuedProp{$s, \smallOP_p$} 
        \State \result.\APPEND  ($\smallOP_p$, \tempzp) \; \label{algo:line:regex-for-value-end}
    \EndFor
    \State $s_p$ = \reForFiniteValuedProp {$s$, \textattrname{Clothes}} \; \label{algo:line:regex-for-clothes}
    \If{$s_p$ is $\phi$}
            $s_p \gets s \setminus$ \tempzp \label{algo:line:no-wearing-for-clothes}
        \EndIf           
        %%%%%%%%%%
        \State \nid $\longleftarrow \emptyset$\   \mysamelinecomment{Index-List for property-name}
        \State \desc $\gets \emptyset$            \mysamelinecomment{List for property-values}
        % \tcc{set of tokens}
        \State \tokens $\leftarrow$ \TOK{$s_p$}    \mysamelinecomment{List of tokens from $s_p$}   % \TOK{$s$}
        % \tcc{set of POS tags}
        \State \tags $\leftarrow$ \POS{\tokens}  \mysamelinecomment{List of $\langle  \text{token, POS-tag} \rangle$ from tokens}
        %
        % \BlankLine
        \mynewlinecomment{\word\ and $\postag_i$ is token and POS-tag at $i^{th}$ index in \tags}
        \For{($w$, \postag) in \tags}
            \If{$\postag_1$ is \emph{VBD}}           \label{algo:line:first_verbs:start}
                \Continue
            \EndIf
            \If{$\postag_2$ is \emph{VBG} and $\postag_1$ is \emph{VBD}}
                \Continue                   \label{algo:line:first_verbs:end}
            \EndIf
            \If{$\postag_{i} \in$ \SetOne $~\cup~$ \SetTwo}   \label{algo:line:property-name-decribe-start} 
                \If{$\postag_{i} ~\in$ \SetTwo\ and $\postag_{i-1} \in$ \SetThree}
                    \State \Break  \label{algo:line:wearing-not-found} 
                \EndIf
                \State \nid.\APPEND (i)\;           \label{algo:line:property-name-decribe-end}
            \ElsIf{$\postag_{i}$ is \adj}           \label{algo:line:adj:start}
                \State \nid $\longleftarrow \emptyset$   \mysamelinecomment{re-initialize name index-list}
                    \State \desc.\APPEND (\word) \;   \label{algo:line:adj:end}
            \ElsIf{$\postag_{i}$ is \Noun}      \label{algo:line:noun:start}
                \State \wm $\gets$ \syn{\word, \Noun}                                 \label{algo:line:noun-color:start} \;
                \State \nid, \desc, \descColor = \newline \null \hfill \fndmtch{\wm, \nid, \desc} \;
                % \lIf{match found with \colour}{\Continue} 
                \If{\descColor}
                    \Continue  \label{algo:line:noun-color:end} 
                \EndIf
                \State \names $\gets$ \word \;      \label{algo:line:multi-name:start}
                \State $\names \gets$ \popopn{\nid, \names, \tags}                               \label{algo:line:multi-name:middle} 
                %
                % \BlankLine
                \mynewlinecomment{finalize property-name \& assign the values}
                \If{$\postag_{i-1}$ is \Noun\ and \newline \null \hfill
                \result[-1].name == $\wordWithoutIdx_{i-1}$} 
                    \State \CONCAT(\result[-1].name, \word, " ")      \label{algo:line:multi-name:end}
                \Else
                    \State \result.\APPEND ($[N, D]$)           \label{algo:line:insert-to-results}
                \EndIf
                %
                %
                % \tcc*{re-initialize Lists of name-index and values}
                \State \nid $\longleftarrow \emptyset$,   
                        \desc $\gets \emptyset$ \mysamelinecomment{re-initialize Lists} \;    
            \Else
                \State \Break
            \EndIf
        \EndFor
\EndFor
% \Return \result \;
\end{algorithmic}
\caption{\posalgo}
\label{algo:cloths-name-value}
\end{algorithm}

%
%
%%%%%%%%%%%%%%%%

%
Algorithm \ref{algo:cloths-name-value} presents the pseudocode of the search technique \posalgo, which takes the sentences in a 
document $T_s$ as input and returns the \textit{collection} of object-properties and their set of values, $\langle\langle o_p, z_p \rangle\rangle$ as output. 
In case of an implicit mention of clothes, we made an assumption that description of \textattrname{clothes} are always followed by descriptions of \textattrname{gender, race}, and/or \textattrname{height}. \\
% Algorithm \ref{algo:functions-in-posid} describes the function definitions in \posalgo.\\
%
\noindent \textbf{Lines \ref{algo:line:cand-start} - \ref{algo:line:cand-end}} Extract the candidate sentences with the \textsc{RE-search}. If results are empty, extract them with semantic or classification models. 
Set of key-phrases $Q_H$ is provided by the system.
% Each system defines its own key-phrases $Q_H$.
\\
\textbf{Lines \ref{algo:line:regex-for-value-start} - \ref{algo:line:regex-for-value-end}}  
Iteratively search for all the finite-valued properties \{\textattrname{gender, race, height}\} in each $C_s$ and append them to output.\\
\texttt{\repropvalues}~ is a regular expression matching function that takes sentence $s$ and property-name $o_p$ as input, and outputs 
\begin{enumerate*}
    \item property-value $z_p$, if $o_p$ is a finite-valued property, or 
    \item partial sentence $s_p$, if $o_p$ is a variable-valued property.
\end{enumerate*}
Each $o_p$ is mapped to a search-string pattern, $s_R$ in $T$.\\
% In Section~ \ref{subsec:attr-extrac-experiment}, we define the object-property and search-string mapping, $T$ for \texttt{\repropvalues}.\\
%
\textbf{Lines \ref{algo:line:regex-for-clothes} - \ref{algo:line:no-wearing-for-clothes}} For \textattrname{clothes}, \texttt{\repropvalues}~ returns 
either a partial sentence $s_p$ starting with \texttt{wearing}, or an empty string. In case of an empty string, keep the remaining string from $L_z$ after discarding the extracted values from lines \ref{algo:line:regex-for-value-start} - \ref{algo:line:regex-for-value-end}.\\
% after the last value from 
% fixed-valued properties. \\
%
\textbf{Lines \ref{algo:line:first_verbs:start} - \ref{algo:line:first_verbs:end}} If first and second token is verb\cbd{in past or present tense in third person singular, or in gerund form}, it is the start for the \textattrname{Relation} property. Following \ref{obs:first_verbs}, ignore consecutive verbs until another tag is encountered. \\
% between the person and the clothes. \\
%
\textbf{Lines~ \ref{algo:line:property-name-decribe-start} - \ref{algo:line:property-name-decribe-end}} 
% Since property names for variable-valued properties like clothes, can span from first mention of VERB to \{DET, CONJ, PREP, PTCP, ADV\} until PRONOUN or no-tag is encountered, capture them as property-name.\\
Following \ref{obs:property-name-decribe}, capture tokens from a VERB until any pronoun or non-tag as a free-form property value for \textattrname{clothes}.\\
%
%
% When candidate sentences do not have the usual pattern for Clothes (Line \ref{algo:line:no-wearing-for-clothes}), 
% cloth names (such as shirt, jeans, etc.) are identified by ignoring pronouns and non-tagged tokens.
%
\textbf{Lines \ref{algo:line:adj:start} - \ref{algo:line:adj:end}} 
% Since adjectives are used for characterizing a property or used with a noun phrase to modify it, adjective phrases in candidate sentences are considered as clothes descriptions.
Capture the adjectives as clothes descriptions, and initialize the next property.
% When an adjective is encountered, the previous property-name has been completed, so re-initialize the name-index list.
\\
\textbf{Lines~\ref{algo:line:noun-color:start}-\ref{algo:line:noun-color:end}} For noun descriptors in the value i.e., grey {\bf{dress}} pants, compare the wordnet-synset meaning for \texttt{color} 
% as noun 
($COLOR_{syn}$) to the noun-token meaning. 
% If the distance between the synset-meaning of a token $w$ and $COLOR_{meaning}$ is greater than the desired threshold, consider $w$ to be a color and add as property-value. % is already described in func.
Since a description is encountered, name-index is re-initialized for the next property-name. \\
% After a color description, ignore looking for other descriptions and advance to the next token. This captures multi-token values i.e., \textit{red and green shirt}. \\
%
\textbf{Lines~\ref{algo:line:multi-name:start}-\ref{algo:line:multi-name:middle}}
If a noun-phrase is not a color, it is considered as cloth-name with multiple tokens i.e., \textit{dress pants, tank top, dark clothing}. Populate the property-name by backtracking the name-index list.\\
% if token in $P_{PAV}$
%
\textbf{Line~\ref{algo:line:insert-to-results}} 
% Perform an additional check for the noun-tokens as property-name to mitigate errors propagated from the part-of-speech tagger. 
If the previous token is NOUN and does not match the last token of the previous property-name, we consider the end of the current property description. Finalize the current property name and value by appending it to the result.
Otherwise, in line~\ref{algo:line:multi-name:end}, amend the last inserted property-name by appending the current token to it.
\paragraph{Generalization}
Algorithm~\ref{algo:cloths-name-value} assumes that the property identifier is intended for human-properties. \posalgo~ can be generalized to any object-properties in the text as long as the property names and type of values are known.
The search string for fixed-valued properties has to be re-designed.
Variable-valued properties following some degree of grammatical structure would be covered by the iterative search pattern in \posalgo. \textsc{Color} will be replaced by the phrase that describes the properties in the corresponding system.
% We tried different query phrases such as entity types, entity names or types of clothes.
$Q_H$'s are highly non-restrictive phrases and can be constructed from entity types or entity names. 

%%%%%%%%%%%%%%%%%%%%%%%%%%%%%%%%%%%%%%%%%%%%%%%%%%%%%%%%%%%%%%%%%%%%%%%%%%%%%%%%%%%

\section{Experiments and Results}

% \textbf{RQ1: Is GED with the predicted features a good alternative for gold annotations?}
% Compare the results with EARS / direct SQL matching with gold annotations with reults of EARS with predicted features.

% \textbf{RQ2: Can FemmIR do this faster/better than EARS and with better performance?}
% For training FemmiR, i) create and save the graphs, ii) save the dictionary.

\label{sec:experiments}
% \input{sections/experiment.tex}
% \subsection{Dataset Construction}
% \label{subsec:dataset-cons}

% hartResultOne
% \begin{table*}
\begin{figure*}[h]
  \centering
  \begin{tabular}{lllllllll}
    \toprule
    %\hline
    & \multicolumn{3}{c}{Attr-Only}  & \multicolumn{3}{c}{Attr-Value}  & $\theta_H$ & $q_H $
                                                    \\ %\cline{2-7}
                                                    %\\ \cmidrule{2-7}
    Models & Precision & Recall & F1-Score           & Precision & Recall & F1-Score  &  &       %\\
                                                    %\\ \hline %\cmidrule(r){2-4} \cmidrule(l){5-7}
                                                    \\ \midrule
    Word2Vec + \posalgo & 0.83 & 0.38 & 0.52 & 0.85 & 0.35 & 0.49 & 0.5 & clothes \\ %\verb|[^.]*clothes[^.]*|
    RE + \posalgo & 0.86 & 0.82 & 0.84 & 0.92 & 0.82 & 0.87 & X & wear \\ %\verb|[^.]*wear[^.]*|
    WordNet + \posalgo & \textbf{0.93} & 0.33 & 0.49 & 0.89 & 0.30 & 0.45 & 0.9 & \textit{clothes} as noun \\
    SBERT + \posalgo & 0.83 & 0.49 & 0.62 & 0.86 & 0.45 & 0.59 & 0.85 & clothes \\
    RE + WordNet + \posalgo & \textbf{0.93} & 0.65 & 0.77 & \textbf{0.92} & \textbf{0.87} & \textbf{0.90} & 0.9 & \textit{clothes} as noun \\
    \textbf{RE + SBERT + \posalgo} & 0.87 & \textbf{0.87} & \textbf{0.87} & \textbf{0.92} & \textbf{0.87} & \textbf{0.90} & 0.85 & clothes \\
    %\multicolumn{3}{c}{NB} & \multicolumn{3}{c}{NA} \\
    \bottomrule
    %\hline
  \end{tabular}
  \caption{Performance of Different Candidate Sentence Extraction Models based on Clothes Property Identification}
  \label{tab:clothes-various-results}
% \end{table*}
\end{figure*}

% \marsResult
% training time & 61 hrs & 72.59 & 67.97 & 59.18 & 76.11 & 67.13 & 79.24 & 70.53 & 79.34 & 71.01 & 44.96 & 0 \\
%%%%%%%%%%
% \begin{table*}[h]
\begin{figure*}[h] 
  \centering
  \begin{tabular}{l |p{0.25in}p{0.25in}| p{0.25in}p{0.25in}| p{0.25in}p{0.25in}| p{0.25in}p{0.25in}| p{0.25in}p{0.25in}| p{0.25in}p{0.25in}} % {l ll ll ll ll ll ll }
  % \begin{tabularx}{0.9\textwidth}{l |ll| ll| ll| ll| ll| ll}
    \toprule
    %\hline
    Properties & \multicolumn{2}{|p{0.25in}|}{CNN (Resnet50)}  & \multicolumn{2}{|p{0.25in}|}{3D-CNN}
    & \multicolumn{2}{|p{0.25in}|}{CNN-RNN} & \multicolumn{2}{|p{0.5in}|}{Temporal Pooling} &
    \multicolumn{2}{|p{0.55in}|}{Temporal Attention} & \multicolumn{2}{|p{0.55in}}{Color Sampling} \\ \cline{2-13} 
    & acc & F1 & acc & F1 & acc & F1 & acc & F1 & acc & F1 & acc & F1           \\ \hline 
    top-color & 75.22 & 73.98 & 67.91 & 65.19 & 70.54 & 67.33 & 74.98 & 73.13 & 76.05 & 74.64 & 44.65 & 38.31 \\
    bottom-color & 73.55 & 54.09 & 59.77 & 36.56 & 67.71 & 44.44 & 71.69 & 47.84 & 70.15 & 46.89 & 45.26 & 15.88 \\
    gender & 90.01 & 89.71 & 86.49 & 76.22 & 90.07 & 89.62 & 91.04 & 90.63 & 91.82 & 91.48 & - & - \\ \hline
    average & \textbf{79.59} & \textbf{72.59} & 67.97 & 59.18 & 76.11 & 67.13 & 79.24 & 70.53 & 79.34 & 71.01 & 44.96 & 27.10 \\
    \bottomrule
  \end{tabular}
  \caption{Comparison of Property Identifiers for Videos with Accuracy (acc) and F1 measure on MARS dataset (\%)}
  \label{tab:marsComparison}
% \end{table*}
\end{figure*}

\paragraph{\textbf{Dataset Construction}} 
% In this section we describe the novel cross-media dataset we built, which consists of incident and investigation reports, news articles, pedestrian videos and image frames. The data were collected from local police departments and university newspapers and are stored in a PostgreSQL database.
For our problem domain, we needed a dataset that did not have similarity or relevance labels but was compatible with existing property identifiers in the literature. During our collaboration with the local police department for the missing person search, we were provided with incident reports and pedestrian videos from the traffic cam. For benchmarking our proposed retrieval methods, we searched for a similar dataset with gold property annotations. 
We adopted
% the \textbf{\datavideo}~ dataset in \cite{solaiman2022computer} and 
the \textbf{MARS} (Motion Analysis and Re-identification Set) person re-identification dataset 
from \cite{zheng2016mars} for the visual modalities. 
% \textbf{\datavideo}~ provides over 20 hours of traffic-cam videos from different locations consisting of \cmt{around} 20 frames each-minute with 3 properties: gender, clothes, and cloth-colors.
MARS consists of 20,478 tracklets from 1,261 people captured by six cameras. There are 16 properties that are labeled for each tracklet,
% in \cite{DBLP:journals/corr/abs-1901-05742}, 
among which we used - \textattrname{gender (\textattrvalue{male, female})}, 
9 \textattrname{bottom-wear colors}, and
% (black, white, red, purple, yellow, gray, blue, green, complex)}, and 
10 \textattrname{top-wear colors}. 
% (black, white, pink, purple, yellow, gray, blue, green, brown, complex)}.
% The primary difference between \datavideo~ and MARS is how the bounding box is defined around each person. For \datavideo, each frame has multiple persons, vehicles and street objects, whereas MARS separates the bounding boxes for each person and forms the tracklets.
For property identifiers in textual modalities, we build a collection of text data, named \textbf{\datatext}~ dataset from newspaper articles, incident reports, press releases, and officer narratives collected from the police department. We scraped local university newspaper articles to search for articles with keywords i.e., \textit{investigation, suspect, `person of interest'} and \textit{`tip line phone number'}. \datatext~ provides ground-truth annotations for 12 properties describing human attributes with the most common being -- \textattrname{gender, 
% (\textattrvalue{male, female}), 
race, 
% (\textattrvalue{Asian, White, Black, Hispanic, Indian}), 
height, clothes 
% (\textattrvalue{shirts, pants, jacket, jeans, t-shirts, docker})
} and \textattrname{cloth descriptions (\textattrvalue{colors})}. %\textit{gender} \texttt{(male, female)}, \textit{race} \texttt{(Asian, White, Black, Hispanic, Indian)}, \textit{height, cloths} \texttt{(shirts, pants, jacket, jeans, t-shirts, docker)} and \textit{cloth descriptions} \texttt{(mostly colors)}.
% TODO: Include Pictures
%, build (medium, skinny, large), facial\_hair (yes, no), tattoos (yes, no), backpack (yes, no), smoking (yes, no), headphones (yes, no). lot more, guiding to git.
Each report, narrative, and press release describes zero, one, or more persons. 
%%%%%%%%%%%%%%%%%%%%%%% deleting the frequency%%%
\cbd{
The frequency of each data type in \datatext~ is: 
\begin{enumerate*}[label=(\roman*)]
    \item newspaper articles: 300,
    \item officer narratives: 40,
    \item press releases: 13,
    \item dispatch reports: 5, and
    \item synthetic narratives: 1500.
\end{enumerate*}
}
%%%%%%%
% Due to privacy reasons, reports need to be redacted before making them public, which is an important challenge in obtaining this kind of data. 

Using the above-mentioned datasets, we built a novel retrieval dataset, \textbf{\mmirdataexp}. %dataset to evaluate the retrieval performance of \system.
\mmirdataexp\ does not rely on explicit similarity labels and based on the gold property annotations, we can identify the relevant data objects for the user. User can specify which properties are important to them and we can tweak the `relevance' label based on that.
The composition statistics for each modality are:
\begin{enumerate}
    \item Image (3270/1100/1144),
    \item Text (296/178/145), and 
    \item Video (1454/499/539)
\end{enumerate}
where (*/*/*) stands for the sizes of training/validation/test subsets.\\
% We describe how we define data similarity in the respective result sections.
%
%
% \noindent Ground Truth.
% Since our multi-modal data has three features in common, we defined the ranking in a scale of exact to approximate match. 
% REMEMBER: We do not know CED at this section
% We ranked the data samples in terms of the number of common properties among them. % with CED. 
For developing the ground truth, we ranked the data samples in ascending order of the mismatched properties. % among them. % 0 penalty highest relevance, then 1, then 2
The properties were chosen depending on the user requirement, and the mismatches were assigned different penalties. In Example~\ref{example:obj-prop-focus-info-need}, 
% we wanted to solve the problem by finding different data sources describing a similar person. 
% For that purpose, 
an officer searches in the following order:
(1) same gender and race, 
(2) same bottom clothing, and
(3) same top clothing.
The intuition behind this is if there is a gender mismatch, they are definitely not the same person. It is possible for a person to change the top clothing in a short span of time, but it is harder to change the bottom clothing. So even if there is a mismatch in top color, there is a chance of it being the same person given a similar time span and vicinity. 
Therefore, we set the penalty for each mismatched property as follows: $rcost$(\textattrname{top-color}) = 1, $rcost$(\textattrname{bottom-color}) = 2, and $rcost$(\textattrname{gender}) = 3, with gender having the highest penalty and hence, the highest importance. 
% Exact matches have an edit cost of zero, so they are the top most in the ranking.
Exact matches are the top most in the ranking with a zero penalty.
%
%%%%%%%%%%%%%%%%%%%%%%%%

%%%%%%%%%%%%%%%%%%%%%%%%%%%%%%%%%%%%%%%%%
\paragraph{\textbf{Settings}} 
%%%%% start-1 %%%%%%%%%
For Word2Vec, we used the 300 dimensional pretrained model from NLTK \cite{nltk2002} trained on 
Google News Dataset\footnote{\href{https://drive.google.com/file/d/0B7XkCwpI5KDYNlNUTTlSS21pQmM/edit}{GoogleNews-vectors-negative300}}. We pruned the model to include the most common words (44K words). 
%% For word and sentence tokenization, we used NLTK's built-in tokenizers. 
From NLTK, we used the built-in tokenizers and the Wordnet package for retrieving the synsets and wu-palmer similarity score.
% ( Wordnet has methods to return word synsets and similarity scores between synsets. )
For SBERT implementation, we used the zero-shot classification pipeline\footnote{\href{https://huggingface.co/transformers/master/main\_classes/pipelines.html\#transformers.ZeroShotClassificationPipeline}{zero-shot-classification}} from transformers package using the SBERT model fine-tuned on Multi-NLI \cite{williams-etal-2018-broad} task. For part-of-speech tagging, we used the averaged perceptron\footnote{\url{https://www.nltk.org/\_modules/nltk/tag/perceptron.html}} 
tagger model. \textit{The manual narratives in the \datatext~ dataset were excluded for property identification task. }
Query phrases used for $C_s$ identification are:\\
% included entity types, entity names, or types of clothes as follows: 
$q_H$ = \{ \texttt{clothes, wear, shirts, pants} \}.\\ % wearing
%
% The regular expressions used for \textbf{RE} model are ($T$ from algorithm):\\
% \textbf{\textit{Race and Gender:}} \hfill 
% \verb|(([\w]+) ([fe||\verb|wo]*ma[n]*[le]*))| \\
% %
% \textbf{\textit{Clothes:}} \hfill 
% \verb|(wear[^.]+)|\\
% %
% \textbf{\textit{Height:}} \hfill \verb|\d+[.]*\d*['’]\s*\d|
%         \verb|+[.]*\d*[\"\”]| \\
% % or   \hfill  \verb|\d+[.]*\d*\s+feet\s+| 
%         % \verb|\d+[.]*\d*\s+|  \verb|inches\s+[tall]*| \\
%
%%%%%%%% end-1 %%%%%%%%%
%%%%%%%% start-2 %%%%%%%%%
We follow the original train/test partition of the MARS \cite{zheng2016mars} dataset for benchmarking the property identifiers in visual modalities. 
% The training set of MARS consists of 8,298 tracklets from 625 people, while the remaining 8,062 tracklets from 626 pedestrians make up the test set while the average number of frames among these tracklets is 60. %The dataset shares no identity in its training and test sets. 
For models in \cite{DBLP:journals/corr/abs-1901-05742, cnn-rnn, 3D-CNN}, we formed a training batch by randomly selecting 32 tracklets, and then by randomly sampling 6 frames from each tracklet.
%to form a training batch, firstly we randomly select K = 32 tracklets from the train set, then n = 6 frames is randomly sampled from each tracklets, so each training batch is formed by K*n frames. 
During testing, 
$F$ frames of each tracklet are randomly split into $\lfloor\frac{F}{n}\rfloor$ groups, and the final \cmt{testing attribute prediction} result is the average prediction result among these groups. % Cross Entropy Loss is chosen as the loss function and Adam with a learning rate 0.0003 is selected as the optimizer in training. 
%%%%%%%%
We used a validation set of mutually exclusive 125 people selected from the training set. 
% differs from \cite{DBLP:journals/corr/abs-1901-05742} in the validation data choice. 
% They validated their model performance on the test set, whereas we selected a validation set from the training set. 
% The validation set consists of 125 people.  % with XXX traclets, double check this number
%%%%%%%%%%%
For color sampling, we used the result from the first frame from each tracklet. 
% to extract the attribute values. % if random is run, mention that
We compared three properties across all models - \textattrname{gender, top color} and \textattrname{bottom color}.
% , except color sampling. 
% Color sampling does not extract \textattrname{gender}.
% For space issue, can delete number of train, test tracklet or how test set is chosen
%%%%%%%% end-2 %%%%%%%%%
%
%%%%%%%%% start - 3 %%%%%%%%%
For the retrieval model, we only considered the synthetically generated part of \datatext. For Munkres, we used the clapper\footnote{\url{https://software.clapper.org/munkres/api/index.html}} API. We did not use the local node-node interaction information from simgnn during the training phase for \system. 
% \system\ different variable values - What would be the node replacement cost and edge replacement cost? 
% - In WLPD, erc = 0, as we assume all features are present, although in reality this is not the case - videos have attributes always missing. 
% *) GCN with degree 1 would have been better, as features are one node away. 
%
%%%%%%%% end-e %%%%%%%%%

% \subsection{Performance of Retrieval Model -- \system}
\paragraph{\textbf{Property Identification in \datatext\ dataset}}
% \label{subsec:attr-extrac-experiment}
%
% Implementation Details
%%%%%%%For regular expression, we used python RE\footnote{https://docs.python.org/3/library/re.html} package. 
%%%%%% https://regexr.com/6s3ge

% \hartResultTwo
% \begin{table}[H]
\begin{table}[h]
    \centering
    % \begin{center}
    \begin{tabular}{llllll}
    \toprule
    \textbf{Attributes} & \textbf{Gender} & \textbf{Race} & \textbf{Height} & \shortstack{\textbf{Clothes}\\Attr-only} & \shortstack{\textbf{Clothes}\\Attr-value} \\ \midrule
    \textbf{Precision} & 0.94  & 0.94 & 0.72 &  0.87  & 0.92 \\%0.93  & 0.92 \\ 
    \textbf{Recall} &  0.73 &  0.73 & 0.57 & 0.87 &  0.87 \\%0.65 &  0.87 \\ 
    \textbf{F1-Score} & 0.82  & 0.82 & 0.63 & 0.87 &  0.90 \\%0.77 &  0.90 \\
    \bottomrule
    \end{tabular}
    % \end{center}
    %\vspace{-0.6cm}
    \caption{Human Attribute Extraction Results} %in \datatext~ Dataset}
    \label{tab:text-attr-perf}
\end{table}
% \end{table}

We compared the baseline RE-model with the other approaches in Section~\ref{subsec:cand-sent-ex}
% using similarity search, classification model, and stacked models 
for finding $C_s$.
% in the \posalgo~ algorithm.
% and then applied the \posalgo~ algorithm from Line~\ref{algo:line:posid-8}. 
Two different set of metrics were used for the evaluation of \textattrname{clothes} identification. (\textbf{Attr-only}) evaluates how efficiently the model identified all clothes, and (\textbf{Attr-value}) calculates the performance of the model in identifying both the attribute and its descriptive values.
% (i.e., \textit{green and red shirt $\rightarrow$ shirt: green, red}). 
For Attr-value, 
%if a cloth is misclassified, it is penalized for both the attribute and all missing descriptions. 
a true positive occurs only when a valid \textit{clothes} name and a correct description of that cloth is discovered. 
% with precision, recall, and F1-measure
%%%%%%
Figure~\ref{tab:clothes-various-results} describes the performance of different candidate sentence extraction models based on the performance of \textattrname{clothes} identification. 
For the baseline, the group of tokens around \texttt{wear} returned three times better F1-score than any other $q_H$.
% i.e., missing, suspect. 
% A regular expression search function will search through the corpus, returning all texts that match the pattern. The corpus can be a single document or a collection. 
With the other models, 
% among all the query phrases 
$q_H$ = \{\texttt{clothes}\} produced the best score. 
(RE + SBERT) stacked model performs best 
% to find accurate attributes and descriptions 
with 87\% and 90\% F1-Scores, for both metrics. Although (RE + Wordnet) has a higher precision score of 93\% for Attr-only, it has a low recall score of only 65\%, indicating over-fitting. 
% Therefore, for better generalization, we use (\textit{RE + SBERT}) for the multi-modal information retrieval.
%%%%%%%%%%%%%%%%%%%%%% Mid Paragraph %%%%%%%%%%%
Based on a property-frequency analysis, we showed the identification results for the most frequent subset of properties in $\bigOP_H$ for \datatext.
Table~\ref{tab:text-attr-perf} shows the performance of \posalgo\ with (RE+SBERT) for stacked model (lines \ref{algo:line:cand-start} - \ref{algo:line:cand-end}).
For gender and race, the model showed the efficacy of the chosen search-pattern with 94\% precision score. 
A recall score of 73\% shows that most people follow similar style for describing gender and race. For \textit{height} with only 57\% recall score, a rule based model is not sufficient due to varied styling. 
%

%%%%%%%%%%%%%%%%%%%%%%%%%%%%%%%%%%%%%%%%%%%%%%%%%%%%%%%%%%%%%%%%%%%%%%%%%%%
\paragraph{\textbf{Property Identifiers in Visual Modalities}}
% \label{subsec:video-results}
%
%
Since MARS has a large amount of ground truths for person attributes, 
% for property identification in videos and images, 
we compared existing models from the \textit{Person Re-Identification} task.
% and bench-marked them for the retrieval task.
%
% \textbf{Compared Models and Evaluation Metrics.} 
From the CNN models, we used the image-based Resnet50 \cite{he2016resnet50} as the baseline. Due to the temporal nature of videos, we also compared the 3D-CNN \cite{3D-CNN}, CNN-RNN \cite{cnn-rnn}, Temporal Pooling and Temporal Attention \cite{DBLP:journals/corr/abs-1901-05742} models. 
As a heuristic-based model, we chose the color-sampling model from \cite{HILDA2020}. 
% For evaluation metrics, we selected the accuracy and F1-score, which is averaged over all the properties.
%
%%%%% Mid-paragraph %%%%
Table~\ref{tab:marsComparison} describes the bench-marking results for the compared models. Resnet50 performed significantly better than other models for bottom-color, while temporal attention worked best for top-color. Considering the average performance on all attributes, we choose the \textit{image-based CNN model} for the retrieval task. Since the properties in our task are all motion-irrelevant, the video-based extraction models do not have a large impact on the performance. In terms of training data and time, color sampling surely has an advantage. Resnet50 needed 513 minutes and the temporal attention model needed 1073 minutes for training, whereas color sampling has zero training time. 
% Without training data, color sampling can be used as a solution for the cold-start problem. But in presence of training data, we would recommend using fine-tuned models.
Color sampling works by isolating body regions and evaluating pixel values, hence the presence of sunlight or clouds may have adversely affected the performance. %\textbf{\texttt{Comment>> Elaborate/ Change explanation}}.
%%%%%%%%%%%%%%%%%%%%%%%%%%%%%%%%%%%%%
%%%%
%%%%%%%%%%
\newcommand{\tabEARsResult}{
    \begin{tabular}{p{0.25in} p{0.25in}  p{0.4in}  p{0.4in} p{0.6in}}
    \toprule
         Query & Target & EARS & \system\ & FGCross-Net \\
    \midrule
         Image & Text & 0.54 & 0.40 & 0.12\\
         & Image & 0.27 & 0.27 & 0.11\\
         & Video & 0.33 & 0.29 & 0.12\\ 
         & All & 0.30 & 0.28 & 0.11\\ \hline
         Text & Text & 1.0 & 0.52 & 0.23\\
         & Image & 0.37 & 0.29 & 0.08\\
         & Video & 0.46 & 0.33 & 0.07\\ 
         & All & 0.43 & 0.31 & 0.10\\ \hline
         Video & Text & 0.62 & 0.43 & 0.09\\
         & Image & 0.30 & 0.29 & 0.11\\
         & Video & 0.37 & 0.30 & 0.31\\ 
         & All & 0.34 & 0.30 & 0.15\\ \hline
         & \textbf{Avg} & \textit {0.44} & \textbf{0.33} & 0.13  \\
    \bottomrule
    \end{tabular}
}

\begin{table}[]
    \centering
    \tabEARsResult
    \label{tab:ears-perf}
    \caption{MAP Performance of \system\ on \mmirdataexp}
\end{table}

\begin{figure}
\centering
  \begin{subfigure}[b]{0.4\textwidth}
    \includegraphics[width=\textwidth]{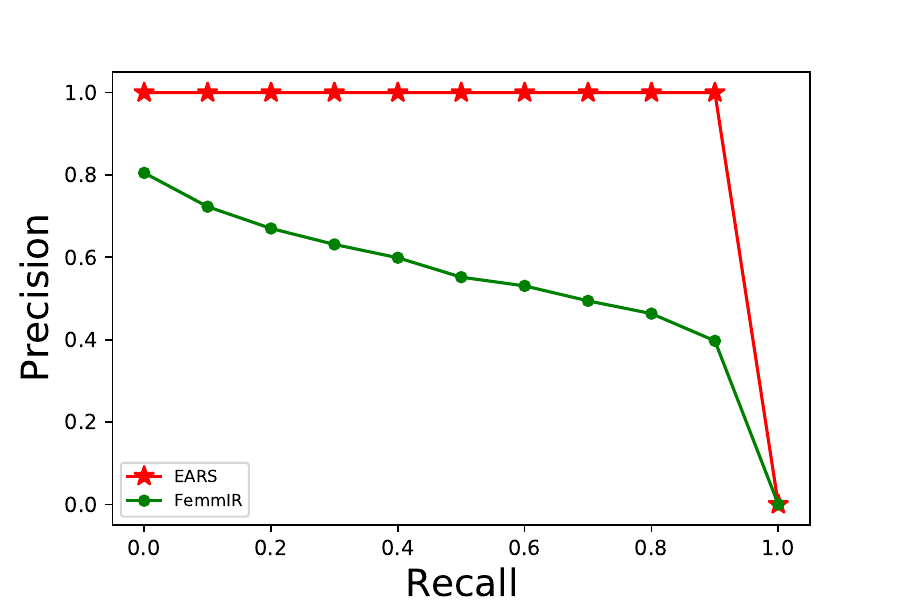}
    \caption{Text $\rightarrow$ Text}
    \label{fig:prcurve-text-query}
  \end{subfigure}
 \hfill
  \begin{subfigure}[b]{0.4\textwidth}
    \includegraphics[width=\textwidth]{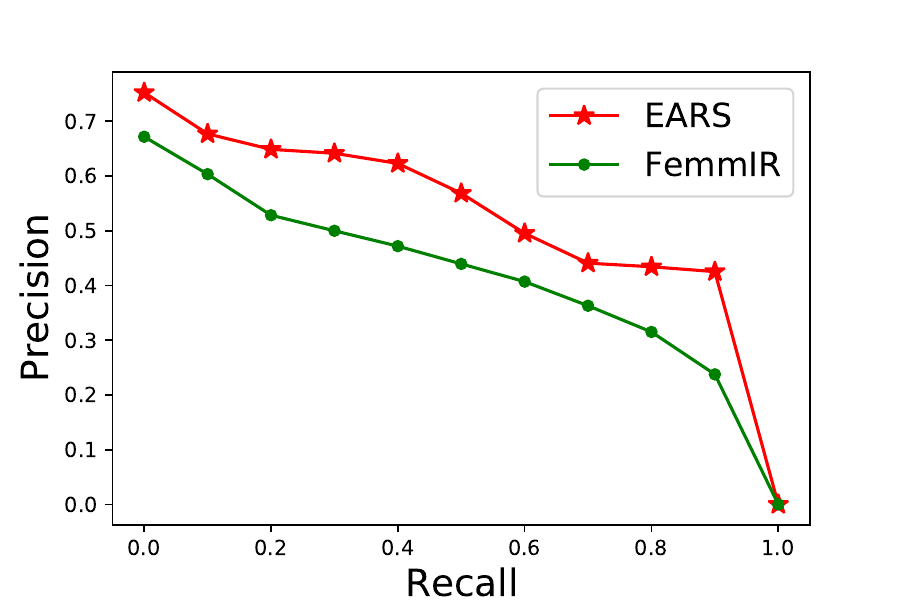}
    \caption{Image $\rightarrow$ Text, Image}
    \label{fig:prcurve-text-retrieval}
  \end{subfigure}
\caption{(a) Precision-recall curves for the text as query and data sample modality, (c) Precision-recall curves for the image as query modality with text and image as data sample modality.}
\label{fig:ears-perf-combined}
\end{figure}

% https://texblog.org/2011/05/24/placing-figures-side-by-side-subfig/
% With minipages you can also create non-nxn-grids, e.g. one on the left, and two stacked on the right:
% \begin{figure}[h]
% \centering
% \begin{minipage}{0.25\textwidth}%
% \subfloat[Subfigure 1 list of figures text]
% [Performance of EARS and \\ \system\ in mAP(\%)]{ %for each modality]{
% \tabEARsResult
% \label{tab:ears-perf}
% } 
% \end{minipage}%
% \begin{minipage}{0.25\textwidth}%
% \subfloat[Subfigure 2 list of figures text][Text $\rightarrow$ Text]{
% \includegraphics[width=\textwidth]{images/graphs_oct15/textQ-textDB.pdf}%{images/text-as-query.pdf}
% \label{fig:prcurve-text-query}
% } \\
% \subfloat[Subfigure 3 list of figures text][Image $\rightarrow$ Text, Image]{
% \includegraphics[width=\textwidth]{images/graphs_oct15/imgQ-textimgDB.pdf}%{images/image-video-as-query-text-as-db.pdf}
% \label{fig:prcurve-text-retrieval}
% }
% \end{minipage}
% \caption{
% Performance of \system\ on \mmirdataexp 
% % : (a) mAP score, (b) Precision-recall curves for the text-query-all-modality, (c) Precision-recall curves for the image-video-query-text-modality.
% }
% \label{fig:ears-perf-combined}
% \end{figure}
%%
%%%%%%%%%%%
\paragraph{\textbf{Retrieval Performance of \system}}
We compared \system\ with EARS \cite{solaiman2022computer} and FGCross-Net \cite{he2019new}. SDML \cite{hu2019deep} has shown superior performance to other state-of-the-art cross-modal retrieval models such as ml-CCA \cite{ranjan2015multi}, ACMR \cite{wang2017adversarial}, GSS-SL \cite{zhang2017generalized}, CMPM+CMPC \cite{zhang2018deep}. Since all of them rely on class labels and our problem setting does not allow fine-tuning, we could only compare the models that provided pre-trained models. SDML does not have any publicly available pre-trained model. EARS is an exact inference model and serves best as a baseline model.  Since EARS does not require any training, we only used the test set in \mmirdataexp. 
We formulated the JOIN queries in the EARS method on the aforementioned properties. The results were a union among the exact match and partial matches for the individual properties. We benchmarked EARS with the extracted property values whereas the gold property annotations were used to create the annotations for multimodal retrieval results.
\cbd{In Example ~\ref{example:obj-prop-focus-info-need}, we have three properties, so a finite number (e.g., three) of partial match queries was needed.}
% Since RelGSim approximates CED calculation, it may result in equal or lower performance than EARS. 
% We are also aware of the state-of-the-art cross-modal retrieval models such as, SDML\cite{hu2019deep}, CMPM+CMPC \cite{zhang2018deep}, and we extracted the pre-trained features for each modality in \mmirdataexp to show comparisons among EARS, RelGSim and SDML. We will include them in an incremental work. %% NO-NO-NO
% We wanted to compare the performance of our multi-modal retrieval model when a good feature extractor exists. Hence for text feature extractor, we used a perfect feature extractor exporting gold annotations as predicted features.
%% NO-NO-NO- we would already have said which identifiers we will be using

%%
%%%%%%%%%%%%%%%%%%%
For evaluation, we considered cross-modal retrieval tasks as retrieving one modality by querying from another modality, such as retrieving text by video query (Video → Text) or, retrieving image by text query (Text → Image). We also show the comparison for multi-modality retrieval. By submitting a query example of any media type, the results of all media types will be retrieved such as (Image → All, Text → All). We adopt mean average precision (mAP) as the evaluation metric, which is calculated on all returned results for a comprehensive evaluation. 
%%%% THe following is what worries me, TODO: recheck the evaluation code %%%%% 
We consider data samples with \alt{CED < 2} $CED < 3$ % maybe in this context I meant this; TODO: recheck
in comparison to the query object, as \textit{\textbf{relevant}} for that query. This would return contents where persons only with color mismatches are found.
% , as per the user requirement.
%%
%%%%%%%%%%

%%%%%%%%%%%%%%%%
% \paragraph{\textbf{Result}}
% Since we used a perfect extractor for text dataset, with query of any modality and target set being text, the mAP score is better than other modalities with (text → text) giving perfect performance, as expected. 
With an average F1 score of 79.59\% for video and image property identifiers, the mAP scores of image and video queries are 27\%-37\%. Text modalities with their high-performance identifiers get the highest mAP across modalities. This indicates \system's correlation with the property identifier performance. If user have a capable property identifier, \system\ performance will increase.
Precision-recall (PR) curves in Figure~\ref{fig:prcurve-text-query} and \ref{fig:prcurve-text-retrieval}
% and Figure~\ref{fig:prcurve-text-retrieval}, 
show that at lower degrees \system\ perform comparably with EARS, but with higher degrees of recall, the performance degrades. \system\ performs significantly better than FGCross-Net without fine-tuning and shows the difficulty of this task.
% We will perform ablation studies (using local node-node interaction or eliminating imbalance of modalities in training data) in the future for \system\ performance improvement.

%%%%%%%%%%%%%%%
%%%%%%%%%%
% Q2. What would be the node replacement cost and edge replacement cost? \\
% - In WLPD, erc = 0, as we assume all features are present, although in reality this is not the case - videos have attributes always missing. \\
% *) GCN with degree 1 would have been better, as features are one node away. \\
% *) During cost matrix calculation or node traversal, leaf nodes are ignored. \\
% Note-3) Different type of node substitution cost = $\infty$\\
% Note-4) Munkers Demo: \href{https://software.clapper.org/munkres/index.html}{link}
%%%  

%%%%%%%%%%%%%%%%%%%%%%%%%%%%%%%%%%%%%%%%%%%%%%%%%%%%%%%%%%%%%%%%%%%%%%%%%%%%%%%%%%%

\section{Related Works}
\noindent \textbf{Metric Learning.} %\cite{liong2016deep, fromedeep, faghri2017vse++, xu2019deep, wei2020universal} learn a distance function over data objects based on a loss function to map multi-modal data in the latent common subspace.
\cite{liong2016deep, fromedeep} uses hinge rank loss to minimize intraclass variation while maximizing interclass variation.
\cite{faghri2017vse++} minimized the loss function using hard negatives with a variant triplet sampling, but needs fine-tuning and augmented data.
\cite{xu2019deep} uses an additional regularization in the loss function with adversarial learning.
%However, these methods treat positive and negative pairs equally.
\cite{wei2020universal} enables different weighting on positive and negative pairs with a polynomial loss function. %assigns a larger weight value to a harder sample. 
\system\ has similarities to metric learning with the objective of minimizing the edit distance between two graphs. In contrast, \system\ re-uses pre-extracted properties and does not require data samples to create positive-negative pairs.
\smallskip 

\noindent \textbf{Weakly Supervised Learning.}
\cite{solaiman2022weakly, mithun2019weakly} use weak signals from entity and relationship similarities retrieved from video captions and text. \cite{solaiman2022computer} assumes knowledge of the translation module which makes it less adaptable to novel modalities. \cite{alaudah2019structureweakly} uses a similarity-based retrieval technique to extract images with similar subsurface structures. \system\ also uses a weak signal approach for ranking relevant samples from multiple modalities, but the weak labels are constrained to use the pre-extracted properties and must implicitly maintain the structure between the entities and relationships. \smallskip

%\noindent \textbf{Multi-modal fusion with Attention Networks and Graph Matching.}
\noindent \textbf{Semantic Understanding with Encoding Networks.}
\cite{lee2018stacked, li2019visual, song2019polysemous, Sah2020AlignedAF} learns semantically enriched representations of multi-modal instances by using global and local attention networks. Similarly, \system\ uses graph convolutional network \cite{kipf2016} to align the most important nodes contributing to the overall similarity, denoting the most similar properties between samples. \smallskip
% \cite{lee2018stacked} proposed a stacked cross attention network for measuring image-text similarity by aligning image regions and words. \cite{li2019visual} used graph convolutional network \cite{kipf2016} to generate relationship-enhanced image region features with a global
% semantic reasoning network for capturing objects and semantic concepts of a scene. \cite{song2019polysemous} computes multiple and diverse representations of an instance by combining global context with locally-guided features via multi-head self-attention and residual learning.
% \cite{Sah2020AlignedAF} uses attention mechanism to align multimodal embeddings learned through a multimodal metric loss function.

\noindent \textbf{Content-based Data Discovery.}
% \cite{LAZARIDIS2013351} employs search and retrieval of multimedia content over databases along with a multimedia indexing scheme by using manifold learning and annotation propagation. \cite{SKOD2019} continuously builds a multi-modal relational knowledge base using SQL queries with extracted features from video and text for delivering data according to user requirements. 
\cite{DICE-elkindi/10.14778/3476311.3476353, solaiman2022computer, Vitrivr/10.1145/3323873.3326921, QbE/Event/10.1145/3397271.3401283, SKOD2019, LAZARIDIS2013351} implement content-based data retrieval by taking user-provided example records as input and returning relevant records that satisfy the user intent. Our work shares similarities to DICE \cite{DICE-elkindi/10.14778/3476311.3476353}, which finds relevant results by finding join paths across tables within the data source.
However, it focuses on discovering relevant SQL queries from user examples, whereas \system\ focuses on finding the relevant content directly by finding similar object properties. EARS \cite{solaiman2022computer} finds relevant data by applying JOIN queries on the user-required properties from different modalities. Similar to EARS, we also assume the knowledge of pre-identified properties. EARS can scale to petabytes of data, but it needs additional queries to retrieve soft similarities. The number of SQL queries increases proportionally to the number of properties in the user query. Contrary to EARS, we do not assume a common schema for all modalities and do not require re-training from scratch to accommodate new modalities.
% \system ~has a similar objective to \cite{konda2016magellan, zhao2019auto} in that it employs entity matching with little or no training data by either fine-tuning or using a pre-trained model, but for data objects. 
\smallskip

% \noindent \textbf{Cross-modal Correlation Learning.}
% \cite{rasiwasia2010CCA, rupnik2010multiMCCA, zhang2017generalized, zhai2013learningwikipedia} use canonical correlation learning %(CCA) 
% to linearly project the low-level features into a common subspace. 
% For non-linear projections, \cite{andrew2013deep, wang2015deep, kan2016multi, peng2017ccl, peng2016cmdn} extended the linear methods  \cite{wang2015deep} or used shallow \cite{peng2017ccl, peng2016cmdn} or deep networks \cite{kan2016multi} to learn the correlations. 
% SDML \cite{hu2019deep} removes the dependency of jointly learning from all modalities by predefining a common subspace 
% and using a deep supervised auto-encoder for each modality.
% % However, since the quantity of information among cross-modal samples is unbalanced and unequal, it is inappropriate to directly match the obtained modality-specific representations across different modalities in a common space. 
% DSRL \cite{Wang2021DRSLDR} directly learns the pairwise similarities by integrating relation learning, capturing the implicit nonlinear distance metric which \system\ also focuses on. Most of these works assume the presence of class labels \cite{zhang2017generalized, zhai2013learningwikipedia}, choice of appropriate feature extraction, and translation models for specific modalities. This limits the capability to integrate new sources or use pre-existing features/properties. \system\ separates the feature extraction modules from the retrieval module and integrates pre-identifier property from any modality using graph encoding networks. \smallskip

Although human attribute recognition from videos and images has been well studied, we believe this is the first work that focuses on finding them from the text.
\cite{chen2018combining, goh2020finding} used sentence encoders and dense neural networks to combine lexical and semantic features for finding similar sentences in electronic medical records and academic writing. 

\section{Conclusion and Future Directions}
\label{sec:conclusion}
We study information retrieval with multimodal (vision and language) queries in real-world applications, which, compared with existing retrieval tasks is more challenging and under-explored.
We introduced the problem of mismatch between the information need and encoder features, along with the lack of annotated data for multi-modal relevance. To this end, we presented \system, a framework that uses weak supervision from a novel distance metric for data objects and uses explicitly mentioned information needs with existing system-identified properties. 
% in intro, you mentioned during exp. what functionality it achieved and data
% here we mention how many models it defeats
% We demonstrated the performance of \system\ in identifying the relevant data to the user example without supervised training and additional resources.
Extensive evaluations
on \mmirdataexp\ dataset demonstrate that
\system\ exhibits strong performance amongst the retrieval models that require fine-tuning and identifies the relevant data to the user example without supervised training and additional resources.
As a byproduct, we also demonstrated the efficacy of HART, a human attribute recognition model from unstructured text, outperforming the baseline language models. \system\ has successfully implemented a \textit{missing person} use case and is being updated to provide further assistance to local agencies in social causes.
In the future, we plan to extend \system\ to include multi-objective and evolving information needs to support more real-world use cases.
%%%%%%%%%%%%%%%%


\section*{Acknowledgment}
This research is supported by the Northup Grumman Mission Systems’ Research in Applications for Learning Machines (REALM) Program.




%\bibliographystyle{IEEEtran}



% \end{thebibliography}

% \bibliography{refs, femmir}
\bibliographystyle{IEEEtran}

\end{document}